\documentclass[aps,amsmath,superscriptaddress,pre,longbibliography,10pt,nofootinbib,citesort]{revtex4-1}

\usepackage{amsmath}
\usepackage{amssymb}

\usepackage[bookmarks,colorlinks]{hyperref}
\usepackage{graphicx}
\usepackage{blindtext}
\usepackage{upgreek}
\usepackage{calc}
\usepackage{amsmath}
\usepackage{amssymb}
\usepackage{bm}
\usepackage{subfigure}
\usepackage[usenames,dvipsnames,svgnames,table]{xcolor}
\usepackage{natbib}

\usepackage{xcolor}

\begin{document}
\title{
Propelled micro-probes in turbulence
}

\author{\textsc{E. Calzavarini}}\email[]{enrico.calzavarini@polytech-lille.fr} 
\affiliation{ \textit{Univ. Lille, Unit\'e de M\'{e}canique de Lille, UML EA 7512, F 59000 Lille, France}}
\author{\textsc{Y. X. Huang}}
\affiliation{ \textit{State Key Laboratory of Marine Environmental Science, College of Ocean and Earth Sciences, Xiamen University,
Xiamen 361102, People's Republic of China}}
\author{\textsc{F. G. Schmitt}}
\affiliation{ \textit{Univ. Lille, CNRS, Univ. Littoral Cote d'Opale, UMR 8187, LOG,Laboratoire d'Oc\'{e}anologie et de G\'{e}oscience, F 62930 Wimereux, France}}
\author{\textsc{L. P. Wang}}
\affiliation{ \textit{UM-SJTU Joint Institute, Shanghai JiaoTong University, Shanghai, 200240, People's Republic of China}}


\date{\today}



\begin{abstract}
%
The temporal statistics  of incompressible fluid velocity and passive scalar fields in developed turbulent conditions is investigated by means of direct numerical simulations along the trajectories of  self-propelled point-like probes drifting in a flow.  Such probes are characterised by a propulsion velocity which is fixed in intensity and direction; however, like vessels in a flow they are continuously deviated by their intended course as the result of local sweeping of the fluid flow. The recorded time-series by these moving probes represent the simplest realisation of  transect measurements in a fluid flow environment. We investigate the non trivial combination of Lagrangian and Eulerian statistical properties displayed by the transect time-series. 
We show that, as a result of the homogeneity and isotropy of the flow, the single-point acceleration statistics of the probes follows a predictable trend at varying the propulsion speed, a feature that is also present in the scalar time-derivative fluctuations. Further, by focusing on two-time statistics we characterize how the Lagrangian-to-Eulerian transition occurs  at increasing the propulsion velocity. The analysis of intermittency of temporal increments  highlights  in a striking way the opposite trends displayed by the fluid velocity and passive scalars. 
\end{abstract}
\maketitle

\section{Introduction}
In the oceanographic and meteorological context, field measurements of fluid velocity and scalar field intensities - such as temperature or concentrations of bio-geo-chemicals - are 
often performed by means of drifting instrumentations either self-propelled or pulled by moving vehicles. We refer for instance to the instrumentation used to perform the very first measurements of turbulence spectra in a open water flow, which was presented in the pioneering paper by Grant \textit{et al.}\cite{Grant_1962}, or to the vast literature on autonomous underwater vehicles (AUV) \cite{WYNN2014451,28797} and their latest miniaturised versions ($\mu$AUV for micro-AUV) \cite{4098869,Watson_2011,Jaffe2017}.  In all the above mentioned cases the probe positions always trace paths across the spatial domain which are the result of the interplay between the direction and intensity of motion imposed by the engine (or by the pulling system)\cite{Osborn1985} and the strength and orientation of the local sweeping of the fluid flow. Such paths are commonly denoted as \textit{transects}. The shape of transect measurements reduces to a rectilinear one if the velocity impressed by the engine is overwhelming compared to the typical fluid velocity while on  the opposite limit, when the engine propulsion velocity is negligible, the transects become the trajectories of ideal fluid tracers.
It is therefore clear that it is only in special limiting cases that transect measurements directly relate to the assessments that can be performed either in a fixed reference system (Eulerian frame) or in the system co-moving with the flow (Lagrangian frame). Because it is natural and convenient to study flows either in the Eulerian or in the Lagrangian frames, there is the need to link the transect measurements to the known and thoroughly studied turbulence phenomenology in such frames.
 
The goal of the present study  is to perform this task in a controlled and idealised setting, \textit{i.e.}, the case of an ensemble of point-like inertia-less propelled probes (and so denoted as microprobes) which drift across a homogeneous and isotropic developed turbulent flow. 
In particular we would like to address the following key questions. How are the fluctuations of the fluid velocity and of a transported scalar quantity varying at changing the probe speed?  Is it possible to grasp the functional dependence of the correlation-time of the recorded signal versus the probe speed? What happens to velocity/scalar time-increments and to their higher statistical moments? In which conditions can we reasonably consider that the probe is sampling a frozen flow/scalar field, which is the classical and often unstated assumption of oceanic cruise measurements? We will see that, despite the simple model of drifting micro-probes under scrutiny, the statistical properties of measured quantities is far from being trivial.

The structure of the paper is as follow. First, we present the theoretical model system used, we illustrate the numerical simulations and the main features of the database obtained by means of a numerical experiment. Second, we describe the data analysis and try, when possible, to substantiate it with analytical or phenomenological predictions and physical interpretation. The focus of the analysis section is both on single and two-point statistics, in particular on fluid  acceleration and velocity and on the scalar concentration field. In the conclusion we discuss the implications of our findings for real micro-probe design and data interpretation and the future perspectives of this research.

 \section{Models and Methods}\label{model}
We consider a model system made of a set of point-like, inertialess, self-propelled probes moving with the following equation of motion: 
\begin{equation}\label{eq:xs}
\dot{\bm{x}}_s(t) = \bm{u}(\bm{x}_s(t), t) + \bm{v}_s
\end{equation}
Here $\bm{u}(\bm{x}_s(t), t)$ denotes the fluid velocity at the position of the probe and  $ \bm{v}_s$ a given time-independent propulsion speed. 
Along these trajectories both the  probe speed $\dot{\bm{x}}_s(t)$  and the local value of a scalar field, $\theta_s(t) \equiv \theta(\bm{x}_s(t),t) $, are recorded. 
The fluid environment in which the probes are placed evolves according to the Eulerian dynamics described by the incompressible Navier-Stokes equation for the velocity field $\bm{u}(\bm{x}(t),t)$ and by the advection-diffusion equation for the scalar $\theta(\bm{x}(t),t)$:
\begin{eqnarray}\label{eq:eul}
\partial_t \bm{u} + (\bm{u}\cdot \bm{\partial}) \bm{u} &=& -\bm{\partial} p + \nu \bm{\partial}^2 \bm{u} + \bm{f}, \quad \ \bm{\partial}\cdot \bm{u} = 0\\
\partial_t \theta + \bm{u}\cdot \bm{\partial}\theta &=& \kappa \bm{\partial}^2 \theta + \Phi
\end{eqnarray}
The force to sustain the turbulent flow, $\bm{f}(\bm{x}(t),t)$, is divergence-less, acts only at large scales, provides a constant global power input, and it is statistically homogeneous and isotropic. The scalar field is characterised by a unit Schmidt number, \textit{i.e.},  the scalar diffusivity, $\kappa$, has the same intensity as the kinematic viscosity, $\nu$, of the advecting fluid. The scalar field is also sustained by a source term, $\Phi(\bm{x}(t),t)$, which acts at large scales and provides a constant global power input (\textit{i.e.} scalar variance) in such a way to obtain a statistically stationary homogeneous and isotropic scalar turbulence. We remark that the scalar field has no feed-back on the fluid flow, it is therefore said to be passive.  The spatial domain is assumed to be tridimensional and cubic, of side $L$, with periodic boundary conditions for all fields. 
In these conditions the turbulent intensity is specified by a single dimensionless parameter, the Taylor-scale based Reynolds number, $Re_{\lambda}$,  while the scalar turbulence is parametrised in terms of a P\'eclet number $Pe_{\lambda_{\theta}}$.\\

The above described model, which is composed by Lagrangian (the probes) and Eulerian (the velocity and scalar fields) elements, can be numerically simulated. In this work the Eulerian dynamics (\ref{eq:eul}) is numerically computed on a Cartesian grid of $N^3$ points through a standard Lattice Boltzmann equation solver based on a single-relaxation time scheme and a double-population algorithm \cite{PhysRevE.55.2780}.  Both the fluid external force and scalar source term are implemented through a combination of low wave-number sinusoidal functions with random phases controlled by independent Ornstein-Ulhenbeck processes, similar to \cite{doi:10.1063/1.4719144} but with overall time dependent amplitudes in order to provide constant mechanical  and scalar power to the system.

The numerically simulated turbulent flows are characterised by the globally averaged numbers $Re_{\lambda} = 75$ and $125$  and correspondingly $Pe_{\lambda_{\theta}} = 43$ and $68$. The values of all the relevant numerical/physical quantities of the simulated flows are reported for completeness in the supplemental material. It is however important to note that in our simulations the spatial dissipative scales, the Kolmogorov scale, $\eta$, as well as the Bachelor scale $\eta_{\theta}$, (which are here equal in magnitude), have been set to $\eta=\eta_{\theta} = 1.5 \delta x$ with $\delta x$ the numerical grid spacing. The time-step of the simulation, $\delta t$, is instead nearly 300 times smaller than the dissipative time-scale $\tau_{\eta}$. This guarantees that the turbulent flow, both for the velocity and the scalar field, is well resolved at all scales (see the supplemental material for a validation test).


The drifting probes evolution equation (\ref{eq:xs}) is integrated in time via $2^{nd}$ order Adams-Bashforth algorithm with the same time stepping, $\delta t$, of the Eulerian algorithm. Trilinear interpolation is used to estimate the values of all the Eulerian quantities at the probe positions. The probe measurements, $\bm{u}$ and $\theta$ together with their gradients, are stored at regular intervals  every $10 \delta t$. 
Drifters have been divided into 21 groups - called probe families -  corresponding to a discrete set of values of their propulsion speed intensity $|\bm{v}_s|$, which is here chosen in the range $\left[ 0 , 9 u_{\eta}\right]$.  There is an extra group of homogeneously distributed and fixed-in-space probes, which we call Eulerian probes. The members of each family differ from each other for the orientation of $\bm{v}_s$, which are taken homogeneously over the solid angle.  This choice has been made just to obtain a faster convergence of the means. Therefore, in this work we denote with $\langle \ldots \rangle$ an average that is both over time and over an ensemble, where the ensembles are isotropic family of probes with equal $|\bm{v}_s|$. 
We track in time a total number of probes $N_p \simeq 2.2\cdot 10^4$ at $Re_{\lambda} = 75$ for a duration in large-eddy turnover units $T_e = L/u_{rms}$ of $\sim 25\ T_e$ and $N_p \simeq 4.4\cdot 10^4$ at $Re_{\lambda} = 125$ for about $13.5\ T_e$. 
Because the observed phenomena at the two different Reynolds number are very similar, in the paper we sometime present only the results from the higher Reynolds and P\'eclet number dataset. However, we will comment whenever necessary on the expected and/or observed Reynolds and P\'eclet dependencies. 

\section{Results}\label{sec:results}
\begin{figure}[!ht]
\begin{center}
\includegraphics[width=1.0\columnwidth]{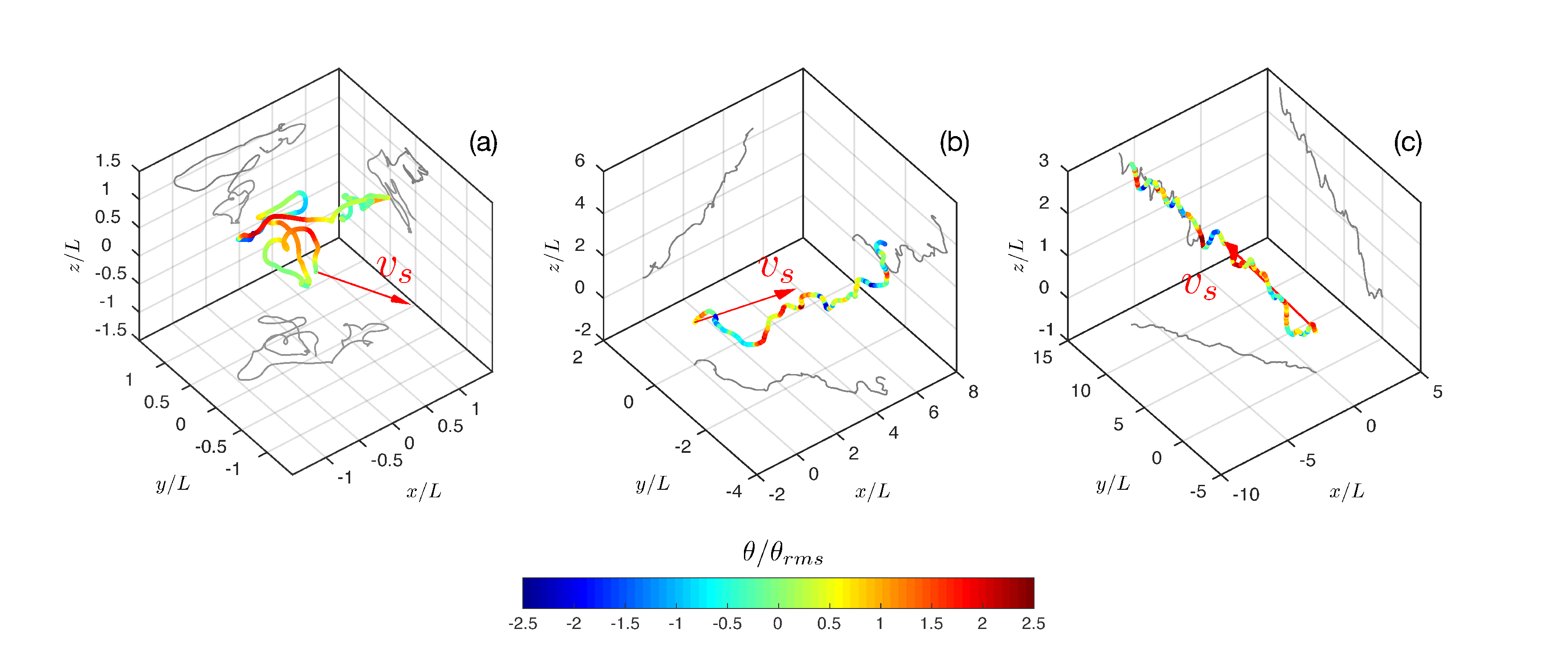}
\caption{Visualization of three drifting probes trajectories with different propulsion velocities in direction and intensity. The velocity intensity is respectively: (a) $v_s=0.9\ u_{\eta}$, (b) $v_s=4.3\ u_{\eta}$ (c) $v_s=9.0\ u_{\eta}$, while the orientation of the propulsion speed $\hat{\bm{v}}_s  = \bm{v}_s / |\bm{v}_s|$ is denoted by the red arrow in each panel. $Re_{\lambda} = 125$ for all cases. The trajectories shown here extends in time over $\sim 8$ large-eddy-turnover time units ($T_e$). The box spatial units are expressed in term of the box size $L$.}\label{fig:visual}
\end{center}
\end{figure}

A visualisation of three typical drifting probe trajectories over a time of $\sim 8T_e$ for different swimming velocity intensities is reported in Fig. \ref{fig:visual}. The shape of the curves (the transects) gets more and more straightened with the increase of the propulsion speed $\bm{v}_s$. From eq. (\ref{eq:xs}) it is indeed expected that a transect will be straight for $v_s \gg u_{rms}$.  At the considered Reynolds number ($Re_{\lambda} = 125$) this corresponds to  $v_s \gg u_{rms} \simeq 10u_{\eta}$. 
From the same figure, which reports distances in term of the box size unit $L$, it is also evident that during the considered timespan the probes have crossed the box spatial domain several times. This is of course allowed by the implementations of periodic boundary conditions, however the box crossing occurs over times-scales long enough for the fluid to loose the past correlations, because $T_e \geq L/\sup{(v_s)}$. 

We now look at the statistical features of the probe measurements. It is easy to verify that in a incompressible fluid flow $\bm{u}(\bm{x}, t)$, as long as $ \bm{v}_s$ is the same for a given set of probes, such probes do not form clusters.  The argument goes as follows, we consider $\dot{\bm{x}}_s(t)$ as a field and take its divergence $\bm{\partial} \cdot \dot{\bm{x}}_s(t) $,  this leads immediately to $\bm{\partial} \cdot \dot{\bm{x}}_s(t) = \bm{\partial}  \cdot  \bm{u} = 0 $ due to the incompressibility of the carrying fluid  flow.  The absence of clustering for the probes has an immediate implication: the statistical moments of the fluid velocity and of the scalar measured by the probes are equal to the Lagrangian $(L)$ and Eulerian $(E)$ averages. Given the absence of the mean flow, \textit{i.e.},   $\langle \bm{u}(\bm{x},t) \rangle_t=0 \ \forall \bm{x}$  with  $\langle \ldots \rangle_t$  denoting the time average, we deduce that  $\langle (\dot{\bm{x}}_s - \bm{v}_s)^n \rangle_t = \langle \bm{u}_L^n \rangle_t  =  \langle \bm{u}_E^n \rangle_t $ for any given even integer  $n$, while all odd-values of $n$ lead to zero moments. Similarly for the scalar, where the mean component $ \langle \theta(\bm{x},t) \rangle_t=0 \ \forall \bm{x}$, one has $ \langle \theta_s^n \rangle_t =  \langle \theta_L^n \rangle_t = \langle \theta_E^n \rangle_t$  for even $n$ and zero for the odd statistical moments.

\subsection{Two time statistics: infinitesimal time gaps} 
As we have observed above the statistical moments of the velocity and scalar recorded by the probes can be easily linked to the Eulerian and Lagrangian values. It is less evident however, how to characterize the behaviour of the moments of velocity and scalar differences in time given for exemple by theirs structure functions or spectra, as classically considered in fully developed turbulence.  We begin this description by considering the limiting case of a difference over an infinitesimal time increment, this is equivalent to the study of the probe acceleration (or the fluid acceleration see by the probe) and of the scalar time derivatives recorded by the probe. 

\subsubsection{Probe acceleration and scalar time-derivative}
\begin{figure}[!htb]
\begin{center}
\includegraphics[width=1.0\columnwidth]{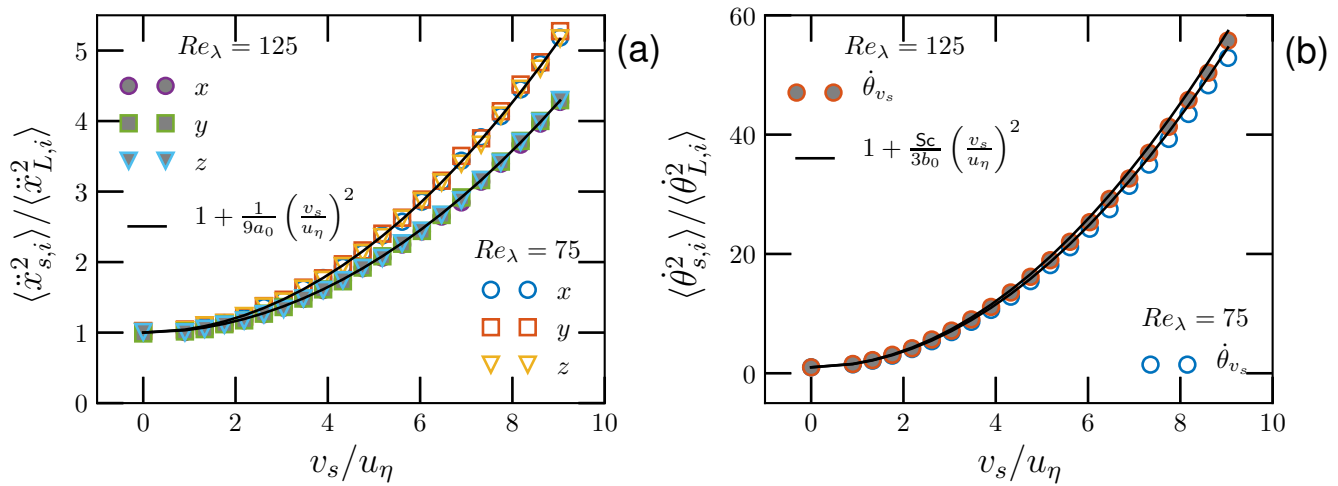}
\caption{a) Variance of the probe acceleration, $\ddot{\bm{x}}_{s,i}$ with index $i$ denoting a Cartesian component, normalized by the variance of the fluid acceleration $\ddot{\bm{x}}_{L,i}$ as a function of the propulsion velocity $v_s$. The continuous line corresponds to the prediction (\ref{eq:acc-prediction}). b) Variance of passive scalar time-derivative and comparison with the prediction (\ref{eq:dT-prediction}).}\label{fig:acceleration}
\end{center}
\end{figure}

The variance of the recorded fluid (or probe) acceleration as well as the scalar time derivative as a function of the propelling speed are reported in figures 
\ref{fig:acceleration} a) and b).
It is evident that both these quantities increase at increasing the propulsion speed.  Furthermore, one can observe that the increase in the variance of the scalar derivative is much larger that the one of the acceleration. What is the origin of such a behaviour and is it possible to account for it?
In order to quantitatively understand this statistical signature, one has to consider that there is a strong similarity between the dynamics of point-like drifting probes and the one of sub-Kolmogorov scale non-neutrally buoyant particles transported by the flow in the presence of gravity. The latter phenomenon has been recently addressed in \cite{Mathai_2016}, where it has been shown that for small inertial particles (bubbles or drops) vertical acceleration fluctuations were lower than horizontal acceleration fluctuations by a factor two, a factor which could be precisely connected to the statistics of local fluid flow gradients in a isotropic turbulent flow. Furthermore, it was observed that the dissipative velocity ($u_{\eta}$) was the relevant scale to the describe the behaviour of the acceleration variance at increasing the particle inertia.
Likewise when one looks at acceleration of a small probe, $\ddot{\bm{x}}_s(t)$, the following relation holds: 
\begin{equation}\label{acc}
\ddot{\bm{x}}_s(t) = D_t \bm{u}( {\bm{x}}_s(t), t)    + \bm{v}_s \cdot \bm{\partial} \bm{u}( {\bm{x}}_s(t), t) 
\end{equation}
where $D_t (\bullet)= \partial_t  (\bullet)+ \bm{u}\cdot \partial (\bullet)$ is the convective derivative along a fluid parcel or tracer. 
In homogeneous and isotropic turbulence it is safe to assume that the velocity gradients are independent  from the material derivative, and along the lines of the derivation given in  \cite{Mathai_2016} one can show (see supplemental material for a detailed derivation), that:
\begin{equation}\label{eq:acc-prediction}
\frac{ \langle \ddot{x}_{s,i}^2 \rangle }{  \langle \ddot{x}_{L,i}^2 \rangle }= 1 + \frac{1}{9 a_0} \left( \frac{v_s}{u_{\eta}}\right)^2
\end{equation}
where  $a_0$ is the so called Heisenberg-Yaglom (HY) constant, a quantity weakly dependent on $Re_{\lambda}$ \cite{voth2002} and defined on the single component of the fluid acceleration $\ddot{x}_{L,i} = D_t u_i(\bm{x}_L(t), t)$ as  $ \langle \ddot{x}_{L,i}^2 \rangle = a_0\ \epsilon/ \tau_{\eta}$ (no summation over $i$ implied).
Therefore, the acceleration variance of a drifting probe is always larger than the one of fluid tracer. The above relation tells also that the higher is the propelling speed of the probe the larger are the acceleration fluctuations experienced by the probe.  The situation is analogous, although at much smaller scale, to the fluctuation due to atmospheric turbulence experienced on airplane, where the intensity of fluctuations depends on its cruising speed \cite{Zbrozek_1958}. Equation (\ref{eq:acc-prediction}) is proven to be well verified on our numerical data, see figure \ref{fig:acceleration} a).

Similarly to the fluid acceleration variance, the  isotropy of the turbulent fluid and scalar flow requires that the time derivative of $\theta_s(t)$ follows:
\begin{equation} \label{eq:dT-prediction}
\frac{ \langle \dot{\theta}_s^2 \rangle }{  \langle \dot{\theta}_L^2 \rangle }= 1 + \frac{Sc}{3 b_0} \left( \frac{v_s}{u_{\eta}}\right)^2
\end{equation}
where  $b_0$ is the scalar equivalent of the HY constant defined as $ \langle \dot{\theta}_L^2 \rangle = b_0\ \epsilon_{\theta}/ \tau_{\eta}$ and $\dot{\theta}_L = D_t \theta(\bm{x}_L(t), t)$ is the time derivative of scalar along a fluid tracer.  Also  eq. (\ref{eq:dT-prediction}) captures very well the numerical simulations, at least for the explored case at $Sc=1$, see fig. \ref{fig:acceleration} b). We also note that for comparable values of $v_s$, the variance of the scalar time derivative, $\dot{\theta}_s$, is much larger than the variance of the fluid acceleration, seen in figure \ref{fig:acceleration} a). By using (\ref{eq:acc-prediction}) and (\ref{eq:dT-prediction}) one can estimate the asymptotical increasing factor as $\sim 3 Sc\ a_0 / b_0$, which we evaluate to be $\sim 13$ at  $Re_{\lambda} = 75$ and $\sim 17$ at  $Re_{\lambda} = 125$. This difference in variances is likely to be originated by the presence of fronts (or ramp and cliffs \cite{Warhaft_2000}) in the spatial structure of the scalar field, sharper than the ones observed in a Cartesian component of the velocity, such fronts are crossed more and more as the probe propulsion is increased and greatly contribute to the scalar fluctuations.

\subsection{Two time statistics: finite time gaps} 
We now look at the behaviour of increments of the probe velocity and of the scalar field across finite-time gaps, and as before we aim at determining the trends as a function of the propulsion velocity. We initially examine the second order statistics through correlation functions and spectra, while later on  higher order moments will be investigated. 

\subsubsection{Correlation function and correlation time}
\begin{figure}[!hb]
\begin{center}
    \includegraphics[width=1.0\columnwidth]{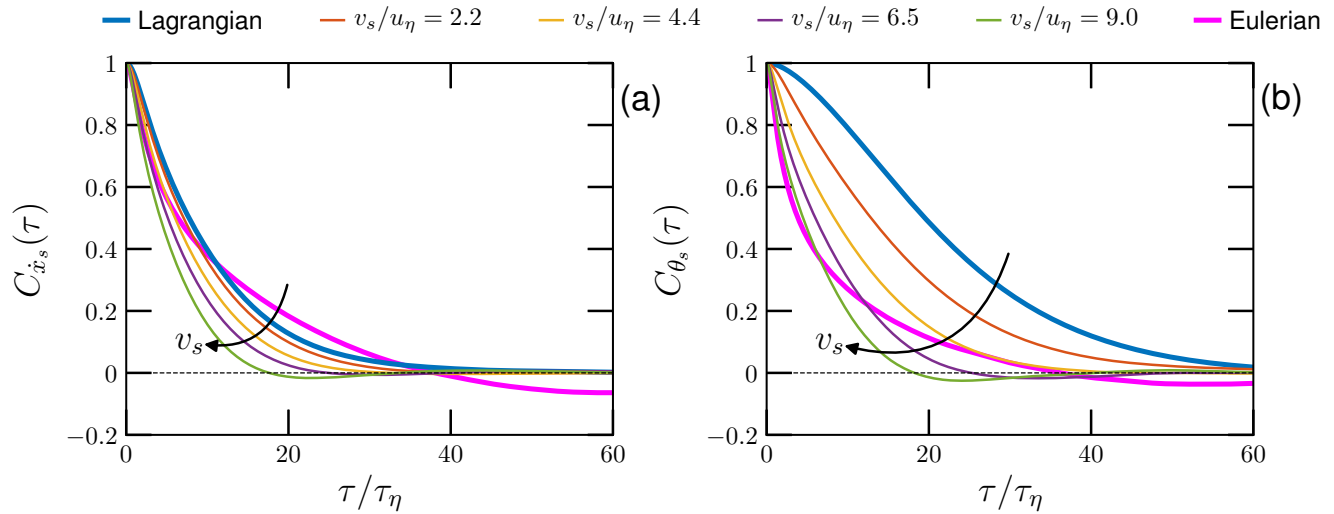}
\caption{(b) Behaviour of the correlation function of the probe velocity $\dot{\bm{x}}_{s,i}$ (single Cartesian component) and (a) Correlation of the scalar value recorded by the probe $\theta_s$ at varying the transect velocity amplitude $v_s$ (the arrows indicate its growing direction).
The correlations functions computed from the fixed-point, denoted as Eulerian, are also reported in both cases.}
\label{correlation-function}
\end{center}
\end{figure}

We consider the  temporal correlation functions of $\dot{x}_{s,i}(t)$ (meaning one Cartesian component) and $\theta_{s}(t)$  at varying $v_s$, whose measurements are reported in figure \ref{correlation-function}. From these graphs it is evident that, as the propulsion velocity is increased, both the velocity and scalar signal become less and less correlated. However, the limiting correlations function at large $v_s$ significantly differs from the one that we measure for fixed-point (Eulerian) probes.
Furthermore, the scalar signal in the Lagrangian frame is much more correlated then the single component of the velocity.

The observed trends in the correlation time can be understood by means of an approximate model.  The starting point is again eq. (\ref{acc}), which is used in the following to build an approximation for velocity increments and later on for the correlations functions. When this equation  is rephrased in terms of small but finite increments, $\tau$ for time and $\bm{v}_s \tau$ for space, it yields :
\begin{eqnarray}
\frac{\dot{\bm{x}}_s(\tau+t) -  \dot{\bm{x}}_s(t)}{\tau} \simeq
 \frac{\bm{u}({\bm{x}}_s(t) + {\bm{x}}_L(\tau), t+\tau) - \bm{u}( {\bm{x}}_s(t), t)}{\tau}  +
    v_{s,j} \frac{\bm{u}( \bm{x}_s(t)+\bm{v}_s \tau, t) - \bm{u}( {\bm{x}}_s(t), t) }{ v_{s,j}\ \tau}
\end{eqnarray}
with summation over $j$ here implied. Note that $\bm{x}_L(\tau)$ denotes here the position at time $t+\tau$ of a fluid tracer that was at the same position of the probe at time $t$.
This simplifies to:
\begin{eqnarray}\label{link}
\delta_{\tau} \dot{\bm{x}}_s \simeq \delta^{L}_{\tau} \bm{u} + \delta^{E}_{\bm{v}_s  \tau} \bm{u}, 
\end{eqnarray}
where we have introduced respectively: i) the temporal increment of the probe velocity, $\delta_{\tau} \dot{\bm{x}}_s = \dot{\bm{x}}_s(\tau+t) -  \dot{\bm{x}}_s(t)$; ii)  the velocity difference along  the trajectory of a fluid tracer, $ \delta^{L}_{\tau} \bm{u}(\bm{x}_s(t) , t) \equiv \bm{u}(\bm{x}_s(t)+ \bm{x}_L(\tau), t+\tau)  -  \bm{u}(\bm{x}_s(t), t) $, which is called Lagrangian velocity difference and iii) the velocity difference across two positions in space at distance $ \bm{l}$ and at a fixed time, $\delta^{E}_{\bm{l}}  \bm{u}(\bm{x}_s(t), t) \equiv \bm{u}(\bm{x}_s(t) + \bm{l}, t) - \bm{u}(\bm{x}_r(t), t)$, which we  call Eulerian difference.
If the  two \textit{rhs} terms in (\ref{link}) are taken as independent, this suggest that in statistically stationary conditions and for a single Cartesian components,  one has:
\begin{eqnarray}\label{eq:sfvel}
\langle  (\delta_{\tau} \dot{x}_{s,i})^2 \rangle(\tau) \simeq \langle  (\delta^{L}_{\tau} u_i)^2 \rangle(\tau) + \langle  (\delta^{E}_{\bm{v}_s  \tau} u_i)^2 \rangle(v_s \tau).
\end{eqnarray}
Note that due to isotropy and homogeneity  the only dependencies left in the above term are respect to time $\tau$ and respect to the amplitude of the displacement  $v_s \tau$.
In the limit $v_s \to 0$ the relation gives the fluid tracer (or Lagrangian) limit while in the opposite asymptote $v_s \to + \infty$ it says that the spatial (Eulerian) increment will dominate. A final caveat should be added concerning the Eulerian increments $\delta^{E}_{\bm{v}_s  \tau} \bm{u} $, which are neither longitudinal or transverse ones but rather a mixture of the two \footnote{The second-order moment of the single-component  Eulerian increment can be rewritten in the form $\langle  (\delta^{E}_{\bm{v}_s  \tau} u_i)^2 \rangle  = \langle (\delta^{E}_{\bm{v}_s \tau \parallel}  u )^2  \rangle + \langle   (\delta^{E}_{\bm{v}_s \tau \perp}  u )^2 \rangle$ if the assumption is made that the longitudinal contribution, $\delta^{E}_{\bm{v}_s \tau \parallel}  u \equiv \delta^{E}_{\bm{v}_s  \tau} \bm{u} \cdot \bm{v}_s/|\bm{v}_s|$  and the transverse one, $ \delta^{E}_{\bm{v}_s \tau \perp}  u \equiv \delta^{E}_{\bm{v}_s  \tau} \bm{u} \times \bm{v}_s/|\bm{v}_s|$,  are statistically independent.}.

\begin{figure}[!htb]
\begin{center}
\includegraphics[width= 1.0\columnwidth]{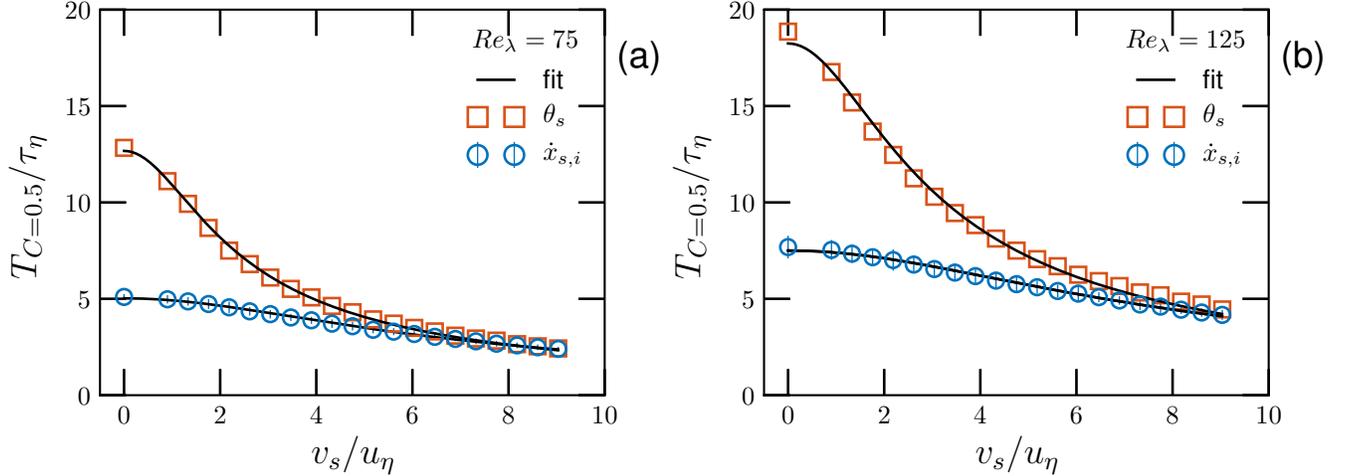}
\caption{Velocity and scalar correlation time, defined as width-at-half-height $T_{C=0.5}$,  at varying the probe drifting velocity intensity $v_s$. Panel (a) reports results at $Re_{\lambda} = 75$, panel (b) at $Re_{\lambda} = 125$. All measurements are well described by eq.(\ref{eq:sqrt})  with fitting parameters $\tilde{\tau}_{\lambda} = 7.1 \tau_{\eta} (10.6 \tau_{\eta})\ $, $\tilde{\lambda} = 34.4 \eta (62.4 \eta)\ $  for the velocity  and   $\tilde{\tau}_{\lambda} = 17.9\tau_{\eta} (25.8\tau_{\eta})\ $,  $\tilde{\lambda} = 30.3\eta (55.3\eta)\ $ for the scalar at $Re_{\lambda} = 75 (125)$.}\label{correlation-length}
\end{center}
\end{figure}

The definition of the correlation function in term of the variance of increments, $C_{\bm{u}}(\tau) = 1 - \langle  (\delta_{\tau} \dot{\bm{x}}_s)^2 \rangle/ (2 \langle \dot{\bm{x}}_s^2 \rangle)$, allows to immediately recast eq. (\ref{eq:sfvel}) in term of correlations: $C_{\dot{\bm{x}}_s}(\tau) \simeq  C^E_{\bm{u}}(\bm{v_s} \tau)  + C^L_{\bm{u}}(\tau) - 1$. 
The correlation time $T_{h}$ for $\dot{\bm{x}}_s$ can be defined as the time at which its correlation function decreases to a value $h$ ($0\leq h < 1$). 
and when the values of $h$ is sufficiently close to 1, one can adopt a quadratic approximation for the Eulerian and Lagrangian correlation functions:
\begin{eqnarray}
C^E_{\bm{u}}(\bm{v_s} \tau)  \simeq  1 - \frac{(\bm{v_s} \tau)^2}{\tilde{ \lambda}^2}, \qquad
C^L_{\bm{u}}(\tau)  \simeq 1 - \frac{\tau^2}{ \tilde{\tau}_{\lambda}^2}
\end{eqnarray} 
The length $\tilde{\lambda}$ and time $\tilde{\tau}_{\lambda}$ are parameters which in the limit of vanishing $\tau$ becomes equal to the 
Taylor micro scale of turbulence ($\lambda$) and to its temporal equivalent in the Lagrangian frame (denoted $\tau_{\lambda}$). Finally, combining these all together we reach the expression for the correlation time that can be tested on the numerical results:
\begin{equation}\label{eq:sqrt}
T_h(v_s) = \sqrt{ \frac{(1-h)  }{\frac{v_s^2}{\tilde{\lambda}^2} + \frac{1}{\tilde{\tau}_{\lambda}^2}} }.
\end{equation}
This correlation time is a decreasing function of $v_s$ which vanishes in the limit of $v_s \to +\infty$, such a limit corresponds however to a finite correlation length $v_s \cdot T_h(v_s)$ by virtue of the Taylor frozen flow hypothesis which becomes fully valid for large $v_s$.
We note that a similar argument as the one proposed for the correlation of the probe velocity can be put forward for the scalar correlation time measured by the probe,
 after assuming that:
$\langle  (\delta_{\tau} \theta_s)^2 \rangle \simeq \langle  (\delta^{L}_{\tau} \theta)^2 \rangle(\tau) + \langle  (\delta^{E}_{\bm{v}_s  \tau} \theta)^2 \rangle (v_s \tau)$. 
Figure \ref{correlation-length} shows the so called width-at-half-height correlation time, $T_{C=0.5}$, which corresponds to taking the value $h=0.5$, in other words the time at which the correlation function has decreased from 1 to the value 1/2.
It is shown that  eq. (\ref{eq:sqrt}) well captures the behaviour of $T_{C=0.5}$ as a function of $v_s$, when the free parameters $\lambda_{u,\theta}$ and $\tau_{\lambda_u , \lambda_{\theta}}$ are tuned. It is remarkable that the value for $\lambda$ is similar in the case of scalar and velocity and it is close to the value of the Taylor scale.
We finally remark that the measurements of the correlation time performed by the fixed point probes appear difficult to be connected to the correlation times evaluated on all the other probes. We suggest that the time-difference measurements performed in such a frame are strongly affected by the large flow structure of the flow and do not lead to representative time-scale for the flow \cite{Biferale_2011}.

\subsubsection{Velocity and scalar energy spectra}
We now look at the energy spectra. Despite the fact that such a quantity is closely related to the correlation function - being just proportional to its Fourier transform -
it allows to better  highlight the scaling behaviour  of the kinetic energy, particularly with respect to the time-frequency $\omega$ in the inertial range of scales $1/T \ll \omega \ll 1/\tau_{\eta}$. 
\begin{figure}[!htb]
\begin{center}
   \includegraphics[width=1.0\columnwidth]{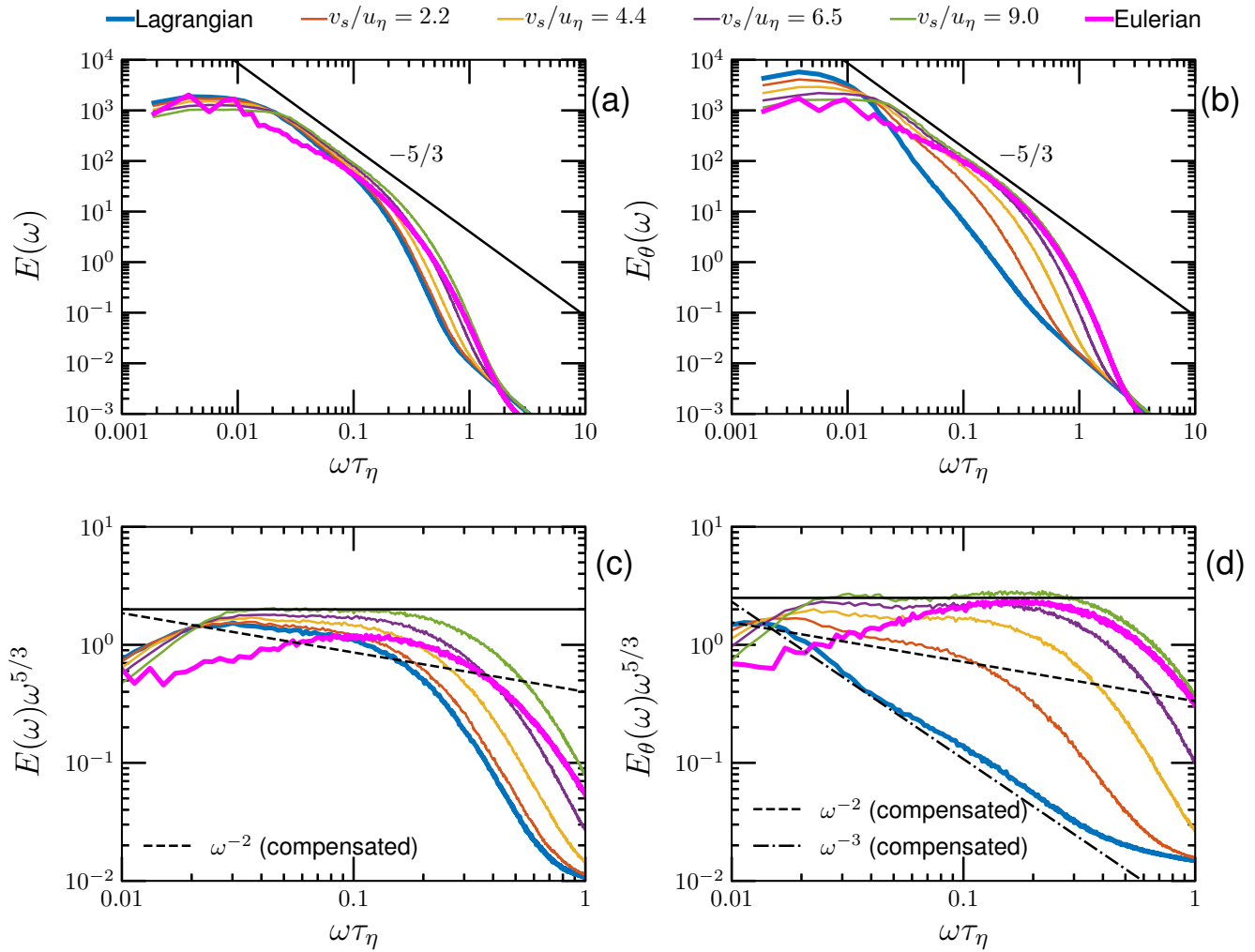}
\caption{Spectra of Kinetic energy (a) and  passive scalar field spectra (b)  at $Re_{\lambda} \simeq 125$.  The compensated spectra with respect to $\omega^{-5/3}$ are reported in panels (c) and (d). 
The power-law scalings $\omega^{-5/3}$, $\omega^{-2}$ , corresponding respectively to turbulence in Eulerian frame or Lagrangian frame and the dissipative scaling and $\omega^{-3}$ are reported.
Note the so called ``flying hot-wire effect'': the progressive shift from Lagrangian to Eulerian scalings at increasing $v_s$: at $v_s \simeq u_{rms}$ the frozen flow approximation becomes fully valid and leads to clean $\omega^{-5/3}$ scalings.}
\label{spectra}
\end{center}
\end{figure}
Indeed, on the basis of dimensional arguments one expects that in the inertial range for the Lagrangian limiting case, 
$E^L_{\bm{u}}(\omega) \sim \omega^{-2}$, while for the Eulerian one, $E^E_{\bm{u}}\left(\omega\right) \sim \omega^{-5/3}$ \cite{tennekes1972first,sh2016book}.  The same is expected for the scalar energy spectra, $E_{\theta}\left(\omega\right)$, in the corresponding inertial range $1/T_{\theta} \ll \omega \ll 1/\tau_{\eta,\theta}$ \cite{tennekes1972first,Sreenivasan96}.
It follows that spectra corresponding to intermediate $v_s$ should fall in-between these two scaling behaviours. Our measurements for a selected number of  probe speeds  are shown in fig. \ref{spectra}. First, we see that the Lagrangian measurements give spectra which are steeper than the Eulerian fixed-probe counterpart. The Lagrangian spectra agrees with similar previous measurements, showing limited scaling ranges \cite{PhysRevE.87.041003}.
 As the probe velocity is increased, both for velocity and scalar, we observe more gentle slopes. The $-5/3$ Kolmogorov scaling becomes particularly evident, in the range $\omega \simeq [0.01,0.1] \tau_{\eta}^{-1}$, for the probes that have a large propulsion velocity.   The progressive shift from temporal-dominate to  spatially-dominated scalings at increasing the probe speed as been long known in the experimental domain and is often indicated as ``flying hot-wire effect'' \cite{Watmuff_1983}.  However it is here remarkable that already for drifting probes at $v_s = 9 u_{\eta} \simeq u_{rms}$  the frozen flow approximation becomes fully valid and leads to a very neat $\omega^{-5/3}$ scalings.  Such a scaling is much more clear than the one observed for the fixed probe. It is indeed known that in turbulent flow that do not possess a mean flow the spectra measured in time domain from a fixed probe (Eulerian measurement) give results different from the expected ones ( \textit{i.e.} from the $-5/3$ scaling). In the condition of no-mean flow indeed the Taylor frozen flow hypothesis can not be applied and the resulting temporal spectra may be affected by the persistence of large scale structures in the flow, as pointed out in \cite{Pinton_1994}. 
We further observe that the scalar spectra for Lagrangian tracers appears to be rather different from the corresponding velocity one. The region compatible with a $\omega^{-2}$ scaling is confined to very large scales, while the rest of the spectra appears to better agree  with a $\omega^{-3}$ behaviour.  We attribute this feature to the limited P\'eclet number of our simulations which are consistent with the numerical results by Yeung \cite{yeung_2001} at similar P\'eclet  and Schmidt values.  The  $\omega^{-3}$ scaling can indeed be obtained by assuming that the scalar variations along a Lagrangian trajectory are smooth, in other words dominated by the diffusion process. This is the counterpart of the observation that the scalar is much more correlated in time as compared to the single component of the velocity field. The steep scaling of the  Lagrangian scalar spectra, estimated  $~ \omega^{-8/3}$,  was also observed  in the mentioned simulations \cite{yeung_2001} and resulted to decrease at increasing the turbulence intensity. Finally, also for the case of scalar measurements we observe that the fixed point spectra does not give consistent results with the $-5/3$ scaling, while the scaling for highly propulsive probes $(v_s \simeq u_{rms})$ are much closer to the expected Obhukhov-Corrsin phenomenology \cite{Sreenivasan96}. 

\subsubsection{High-order statistical moments of velocity and scalar increments}
\begin{figure}[!htb]
\begin{center}
  \includegraphics[width=1.0\columnwidth]{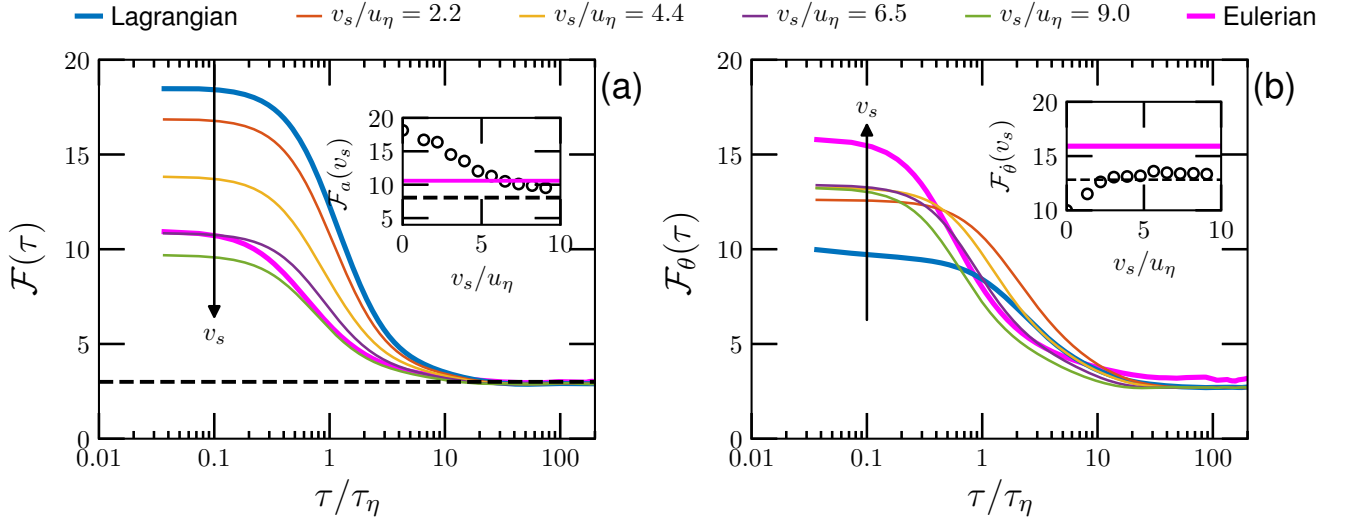}
\caption{(a)Flatness of probe velocity time increments $\mathcal{F}_{\delta_{\tau}\dot{x}_{s,i}}$  as a function of the increment $\tau$ for different probes velocity amplitudes $v_s$. The horizontal dashed line indicates the flatness value of a Gaussian random variable $\mathcal{F} = 3$. In the inset the behaviour of the same quantity in the vanishing time-gap limit, $\mathcal{F}_{\delta_{\tau}\ddot{x}_{s,i}}$  is shown. The horizontal dashed line shows the asymptotic ($v_s \to +\infty$) prediction  $\mathcal{F}_{\delta_{\tau}\ddot{x}_{s,i}} = 3/25\ \mathcal{F}(\partial_{\parallel} u_i) +   24/25\ \mathcal{F}(\partial_{\perp} u_i) $.
(b) Flatness of scalar time increments $\mathcal{F}_{\delta_{\tau}\theta_{s}}$. The inset shows the behaviour of the same quantity in the vanishing time-gap limit, $\mathcal{F}_{\delta_{\tau}\dot{\theta}_{s}}$. The dashed line gives the expected asymptotic limit $\mathcal{F}_{\delta_{\tau}\dot{\theta}_{s}} =  \mathcal{F}_{\partial \theta}$. 
In both panels the arrows denote the trend at increasing the intensity of the propulsion velocity, $v_s$.}
\label{fig:flatness}
\end{center}
\end{figure}

How are higher statistical moments of velocity and scalar time differences affected by the propulsion velocity?\\ 
In order to clarify this point we focus on the flatness of such quantities as a function of the time-gap $\tau$:
\begin{equation}\label{eq:flatness}
\mathcal{F}(\tau) = \frac{ {\langle  (\delta_{\tau} \dot{\bm{x}}_s)^4 \rangle} }{ {\langle  (\delta_{\tau} \dot{\bm{x}}_s)^2 \rangle^2}}\quad , \quad
\mathcal{F}_{\theta}(\tau) = \frac{ {\langle  (\delta_{\tau} \theta)^4 \rangle} }{ {\langle  (\delta_{\tau} \theta)^2 \rangle^2}},
\end{equation}
and we compute them for different propulsion velocities. We begin by looking at the behaviour of the velocity field, Fig. \ref{fig:flatness} (a). For small gaps and zero propulsion velocity the flatness is at its maximum (it is about $~18$ at $Re_{\lambda} =125$).  
We note that the vanishing time-gap limit corresponds to the flatness of the probe acceleration, $\mathcal{F}_{\delta_{\tau}\ddot{x}_{s,i}}$,  which is shown in the inset of the same figure. 
The latter quantity for increasing $v_s$ sample mostly the statistics of the velocity field, and in the asymptotic limit ($v_s \to +\infty$) one can estimate it will reach the plateau  $\mathcal{F}_{\delta_{\tau}\ddot{x}_{s,i}} = 3/25\ \mathcal{F}(\partial_{\parallel} u_i) +   24/25\ \mathcal{F}(\partial_{\perp} u_i) $ (demonstration in supplemental material), a prediction which is consistent with our simulations (see again the inset).
When the time-gaps $\tau$ are increased, all the curves monotonically decrease to the Gaussian value ($\mathcal{F}=3$). Similarly to the trends observed for the limit $\tau \to 0$, we can also observe that the increase of the propulsion velocities $v_s$ produce a progressive reduction in the flatness. For the largest velocities we measure flatness values that at all time-gap $\tau$ are close to the ones that are detected in the Eulerian fixed-point system, meaning probably that relative higher order statistics are less affected by the spurious effects introduced by the absence of a mean-flow.

We now turn the attention on the scalar time-increments flatness,  Fig. \ref{fig:flatness} (b).  In the present analysis it appears that the most intermittent situation occurs again for small time gaps, but this time for the highly movable probes ($v_s = 9 u_{\eta} \simeq u_{rms}$). This is confirmed by the inset, where the vanishing time limit is shown $\mathcal{F}_{\delta_{\tau}\dot{\theta}_{s}}$, together with the asymptotic  ($v_s \to +\infty$)  prediction $\mathcal{F}_{\delta_{\tau}\dot{\theta}_{s}} =  \mathcal{F}_{\partial \theta}$. Contrary to what happens to the velocity we note that we obtain here a reduced flatnesses for decreasing values of the probe speed, with a minimum occurring in the Lagrangian case (note the arrows directions in Fig. \ref{fig:flatness}a) and b)) . The low intermittent level of Lagrangian scalar time increments, has been already observed by Bec \textit{et al.} \cite{PhysRevLett.112.234503}. In the large gap limit we do not exactly find the gaussian flatness value for the probes, but rather a weakly sub-Gaussian one ($\sim 2.7$). This points to long range correlations which seem to be non detectable in the fixed-position probe system. Finally, note that in all cases the scalar is less intermittent than the velocity field.

\begin{figure}[!htb]
\begin{center}
  \includegraphics[width=1.0\columnwidth]{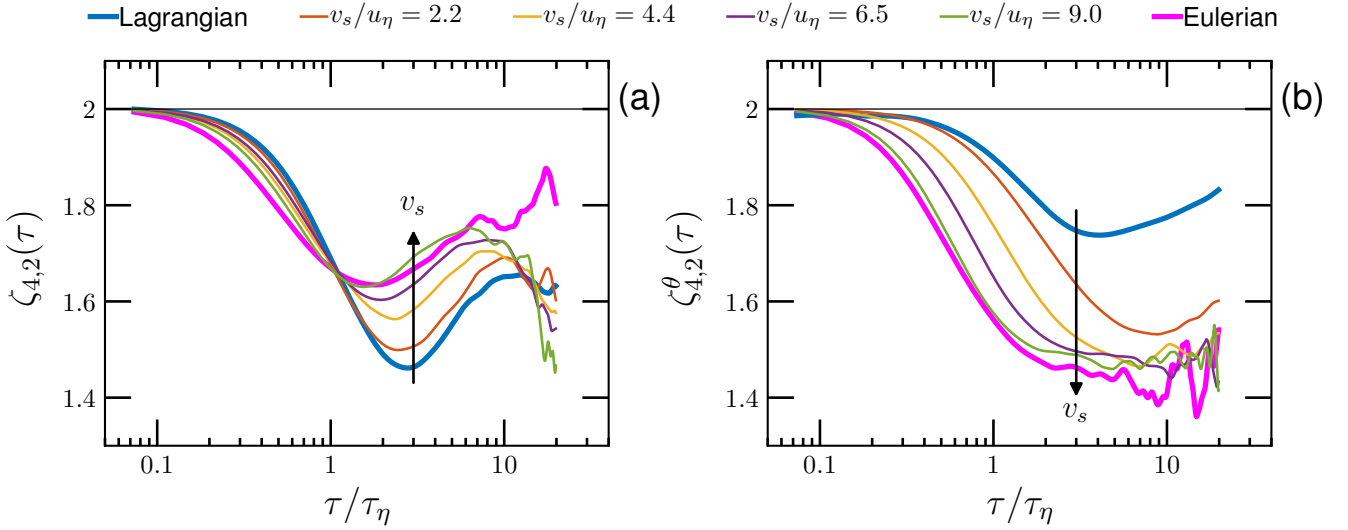}
\caption{(a) ESS local slopes, \textit{i.e.}, logarithmic derivative of the fourth order moments of probe velocity time increments versus the second order one as in eq. (\ref{eq:zeta}). (b) Same as before for scalar increments. Remark the opposite intermittency trends between the velocity and the scalar at increasing $v_s$ intensity (arrow direction).}
\label{fig:zeta}
\end{center}
\end{figure}

To have a deeper understanding on the scaling behaviour as a function of the time gap, in Figure \ref{fig:zeta}, we plot the local slopes by using Extended Self Similarity (ESS), \textit{i.e.}, the logarithmic derivative of the fourth order  temporal structure function, $\langle  (\delta_{\tau} \dot{\bm{x}}_s)^4 \rangle$, versus the second order one, $\langle  (\delta_{\tau} \dot{\bm{x}}_s)^2 \rangle$, which gives  \cite{Arneodo_2008}:
\begin{equation}\label{eq:zeta}
\zeta_{4,2}(\tau) = \frac{ d\log{\langle  (\delta_{\tau} \dot{\bm{x}}_s)^4 \rangle} }{ d\log{\langle  (\delta_{\tau} \dot{\bm{x}}_s)^2 \rangle}} 
\quad , \quad 
\zeta^{\theta}_{4,2}(\tau) = \frac{ d\log{\langle  (\delta_{\tau} \theta)^4 \rangle} }{ d\log{\langle  (\delta_{\tau} \theta)^2 \rangle}}
\end{equation}
The above quantity is a direct scale-by-scale measurement of the local scaling properties. A scale-independent behaviour of the fourth moment against the second-order one would result in a constant value for the left hand side of (\ref{eq:zeta}). Furthermore, in the absence of intermittency, these curves would be constant across the time lags with $\zeta_{4,2}(\tau)=\zeta^{\theta}_{4,2}(\tau) = 2$. While the latter relation is always well verified for the smooth dissipative scales ($\tau < 0.1 \tau_{\eta}$), a non trivial behaviour appears in the inertial range pointing out the intermittent feature of the turbulent flow.  In the time range from 1 to 10 in $\tau_{\eta}$ units, the strong deviation observed in the local slope for the velocity have been attributed to events of tracer trapping in intense vortex filaments \cite{Biferale2005}. We observe that these events tend to fade out as soon as $v_s$ is increased, indicating that fast enough probes can escape from such filaments. However, it is interesting to observe that the deepening associated to this highly intermittent range seems to remain for the case of the highest propulsion velocities and to be close, if not coincident, to the level of intermittency measured in the Eulerian fixed-point probe. This observation questions the vortex-filament interpretation of the so called bottleneck effect \cite{Buzzicotti2016}.
The behaviour of the scalar, Fig.  \ref{fig:zeta} b), deserves further attention. First, we observe that the scalar is highly intermittent both in the fixed-position frame and in the fast drifting one. The values of the exponent we detect here is as low as  $1.5 \pm 0.1$ and this all the way above time gaps $\tau \gtrsim 1 \tau_{\eta}$. Such a strongly intermittent behaviour has been known for a long time \cite{Sreenivasan1561} and it has been associated with the presence of scalar sharp fronts. 
At decreasing the propulsion velocity we observe much less intermittent fluctuations till reaching, at $v_s =0$, the Lagrangian case that by far the less intermittent case  (with $ \zeta^{\theta}_{4,2}(\tau) \sim 1.8$), although a clear inertial range plateau can not be attained in the present simulations.

\section{Conclusions}
We have performed a numerical study of the statistical properties of fluid velocity and scalar concentration fields as measured by small ideal propelled probes in a turbulent flow environment. 
It has been highlighted that despite the simplicity of the model system taken into consideration, the properties of the recorded signals are far from being trivial because they mix Lagrangian and Eulerian turbulent statistical properties. We have first focused on time derivatives of velocity (so acceleration) and scalar concentrations. It appears that, as the propulsion velocity is increased, the probe samples more and more the spatial gradient properties of the fields and we have predicted that such a behaviour evolves quadratically with the ratio of the propulsion speed intensity over the Kolmogorov velocity $v_s/u_{\eta}$, with a weak Reynolds, P\'eclet number dependencies parametrised by the Heisenberg-Yaglom coefficient, $a_0$, and its scalar equivalent, $b_0$.  Such an increase of variance with the propulsion speed is remarkably larger for the scalar as compared to the fluid velocity field, and it signals the higher spatial variations of the scalar field. 
Our analysis of finite time-increments of fluid velocity and of the scalar  encompasses the correlation functions and correlation time, for which an empirical model has been proposed, and the Fourier spectra. It is worth noticing that this analysis demonstrates that probes with a propulsive speed of the order of the turbulent root-mean-square velocity fluctuation, $v_s \simeq u_{rms}$, already display nearly straight line trajectories and can be considered as good tools to measure the Eulerian properties of the flow. Indeed measurements in such conditions reveals the expected $-5/3$ scaling both for velocity and for the scalar in a much clearer way than for the fixed-point measurements where, due to the absence of mean flow, the Taylor frozen-flow hypothesis fails. The two-point higher order statistics highlight in a striking way the opposite trends displayed by the velocity and scalar fields. While the velocity field  in the Lagrangian frame is characterised by wide fluctuations and so by stronger intermittent scaling properties as compared to the Eulerian one, the reverse is true for the scalar. In our opinion it remains to be verified if the origin of such differences is originated by the presence of rather different coherent structures, namely vortex filaments for the fluid and ramp-cliff fronts for the scalar. 

Finally, we would like to readdress the applicative questions that originally motivated this study: The consequence for $AUV$ or boats drifting in turbulent ocean. From the above discussion it is clear that two relevant velocity scales are at play: $u_{\eta}$ and $u_{rms}$. While the former has a role in the description of the time-derivatives statistics of the measured fields, the latter is important for  finite-time gaps observables and spectra. Such scales in the upper ocean have been estimated to be of the order $\mathcal{O}(1)\ cm/s$ and $\mathcal{O}(10)\ cm/s$ respectively \cite{Jimenez_1997,Thorpe2007book,KOROTENKO201321}.  This implies that the most common $AUV$ systems, which typically propel at much larger velocities, $\mathcal{O}(1)$ m/s \cite{Thorpe2007book},  will effectively probe frozen flow turbulent properties (\textit{e.g.} for spectra or structure functions) while miniaturised versions, $\mu AUV$, which have less power and so smaller propulsive force, might find themselves in transitional Lagrangian/Eulerian measurements regime. 
However, the probe model system adopted in our study is still rather simplistic as it neglects, among other effects, the probe inertia, the hydrodynamic drag and the buoyancy force.  The shape of the probe may also have a crucial role, since it can affect the hydrodynamic torque due to the surrounding flow.  These are still open points which need to be addressed in future works.  We conclude by observing that this study is relevant not only for the domain of measurements in the ocean, but it can also be suitable for the description of the perceived fields by small swimming organisms attracted by a given point or direction in space.  Indeed, similar modeling have been adopted for recent studies of swimming phototactic algae or bacteria in turbulent \cite{Torney_2008} or chaotic flow environments \cite{Dervaux_2017}.\\

\textit{Acknowledgments:} The present study has been supported by the Sino-French (NSFC-CNRS) joint Project (No. 11611130099, NSFC China) - (PRC 2016-2018 LATUMAR \textit{Lagrangian turbulence: numerical studies and marine experimental applications}, CNRS France). Y.H. is partially  supported by the National Natural Science Foundation of China (Nos. 11332006 and 11732010).


\begin{thebibliography}{31}%
\makeatletter
\providecommand \@ifxundefined [1]{%
 \@ifx{#1\undefined}
}%
\providecommand \@ifnum [1]{%
 \ifnum #1\expandafter \@firstoftwo
 \else \expandafter \@secondoftwo
 \fi
}%
\providecommand \@ifx [1]{%
 \ifx #1\expandafter \@firstoftwo
 \else \expandafter \@secondoftwo
 \fi
}%
\providecommand \natexlab [1]{#1}%
\providecommand \enquote  [1]{``#1''}%
\providecommand \bibnamefont  [1]{#1}%
\providecommand \bibfnamefont [1]{#1}%
\providecommand \citenamefont [1]{#1}%
\providecommand \href@noop [0]{\@secondoftwo}%
\providecommand \href [0]{\begingroup \@sanitize@url \@href}%
\providecommand \@href[1]{\@@startlink{#1}\@@href}%
\providecommand \@@href[1]{\endgroup#1\@@endlink}%
\providecommand \@sanitize@url [0]{\catcode `\\12\catcode `\$12\catcode
  `\&12\catcode `\#12\catcode `\^12\catcode `\_12\catcode `\%12\relax}%
\providecommand \@@startlink[1]{}%
\providecommand \@@endlink[0]{}%
\providecommand \url  [0]{\begingroup\@sanitize@url \@url }%
\providecommand \@url [1]{\endgroup\@href {#1}{\urlprefix }}%
\providecommand \urlprefix  [0]{URL }%
\providecommand \Eprint [0]{\href }%
\providecommand \doibase [0]{http://dx.doi.org/}%
\providecommand \selectlanguage [0]{\@gobble}%
\providecommand \bibinfo  [0]{\@secondoftwo}%
\providecommand \bibfield  [0]{\@secondoftwo}%
\providecommand \translation [1]{[#1]}%
\providecommand \BibitemOpen [0]{}%
\providecommand \bibitemStop [0]{}%
\providecommand \bibitemNoStop [0]{.\EOS\space}%
\providecommand \EOS [0]{\spacefactor3000\relax}%
\providecommand \BibitemShut  [1]{\csname bibitem#1\endcsname}%
\let\auto@bib@innerbib\@empty
\bibitem [{\citenamefont {Grant}\ \emph {et~al.}(1962)\citenamefont {Grant},
  \citenamefont {Stewart},\ and\ \citenamefont {Moilliet}}]{Grant_1962}%
  \BibitemOpen
  \bibfield  {author} {\bibinfo {author} {\bibfnamefont {H.~L.}\ \bibnamefont
  {Grant}}, \bibinfo {author} {\bibfnamefont {R.~W.}\ \bibnamefont {Stewart}},
  \ and\ \bibinfo {author} {\bibfnamefont {A.}~\bibnamefont {Moilliet}},\
  }\bibfield  {title} {\enquote {\bibinfo {title} {Turbulence spectra from a
  tidal channel},}\ }\href@noop {} {\bibfield  {journal} {\bibinfo  {journal}
  {J. Fluid Mech.}\ }\textbf {\bibinfo {volume} {2}},\ \bibinfo {pages} {241--
  268} (\bibinfo {year} {1962})}\BibitemShut {NoStop}%
\bibitem [{\citenamefont {Wynn}\ \emph {et~al.}(2014)\citenamefont {Wynn},
  \citenamefont {Huvenne}, \citenamefont {Bas}, \citenamefont {Murton},
  \citenamefont {Connelly}, \citenamefont {Bett}, \citenamefont {Ruhl},
  \citenamefont {Morris}, \citenamefont {Peakall}, \citenamefont {Parsons},
  \citenamefont {Sumner}, \citenamefont {Darby}, \citenamefont {Dorrell},\ and\
  \citenamefont {Hunt}}]{WYNN2014451}%
  \BibitemOpen
  \bibfield  {author} {\bibinfo {author} {\bibfnamefont {R.~B.}\ \bibnamefont
  {Wynn}}, \bibinfo {author} {\bibfnamefont {V.~A.I.}\ \bibnamefont {Huvenne}},
  \bibinfo {author} {\bibfnamefont {T.~P.~Le}\ \bibnamefont {Bas}}, \bibinfo
  {author} {\bibfnamefont {B.~J.}\ \bibnamefont {Murton}}, \bibinfo {author}
  {\bibfnamefont {D.~P.}\ \bibnamefont {Connelly}}, \bibinfo {author}
  {\bibfnamefont {B.~J.}\ \bibnamefont {Bett}}, \bibinfo {author}
  {\bibfnamefont {H.~A.}\ \bibnamefont {Ruhl}}, \bibinfo {author}
  {\bibfnamefont {K.~J.}\ \bibnamefont {Morris}}, \bibinfo {author}
  {\bibfnamefont {J.}~\bibnamefont {Peakall}}, \bibinfo {author} {\bibfnamefont
  {D.~R.}\ \bibnamefont {Parsons}}, \bibinfo {author} {\bibfnamefont {E.~J.}\
  \bibnamefont {Sumner}}, \bibinfo {author} {\bibfnamefont {S.~E.}\
  \bibnamefont {Darby}}, \bibinfo {author} {\bibfnamefont {R.~M.}\ \bibnamefont
  {Dorrell}}, \ and\ \bibinfo {author} {\bibfnamefont {J.~E.}\ \bibnamefont
  {Hunt}},\ }\bibfield  {title} {\enquote {\bibinfo {title} {Autonomous
  underwater vehicles (AUVs): Their past, present and future contributions to
  the advancement of marine geoscience},}\ }\href {\doibase
  https://doi.org/10.1016/j.margeo.2014.03.012} {\bibfield  {journal} {\bibinfo
   {journal} {Marine Geology}\ }\textbf {\bibinfo {volume} {352}},\ \bibinfo
  {pages} {451 -- 468} (\bibinfo {year} {2014})}\BibitemShut {NoStop}%
\bibitem [{\citenamefont {Perry}\ and\ \citenamefont {Rudnick}(2003)}]{28797}%
  \BibitemOpen
  \bibfield  {author} {\bibinfo {author} {\bibfnamefont {M.~J.}\ \bibnamefont
  {Perry}}\ and\ \bibinfo {author} {\bibfnamefont {D.~L.}\ \bibnamefont
  {Rudnick}},\ }\bibfield  {title} {\enquote {\bibinfo {title} {Observing the
  ocean with autonomous and lagrangian platforms and sensors: The role of alps
  in sustained ocean observing systems},}\ }\href {\doibase
  10.5670/oceanog.2003.06} {\bibfield  {journal} {\bibinfo  {journal}
  {Oceanography}\ }\textbf {\bibinfo {volume} {16}},\ \bibinfo {pages} {31--36}
  (\bibinfo {year} {2003})}\BibitemShut {NoStop}%
\bibitem [{\citenamefont {Walker}(2006)}]{4098869}%
  \BibitemOpen
  \bibfield  {author} {\bibinfo {author} {\bibfnamefont {D.}~\bibnamefont
  {Walker}},\ }\bibfield  {title} {\enquote {\bibinfo {title} {Micro autonomous
  underwater vehicle concept for distributed data collection},}\ }in\ \href
  {\doibase 10.1109/OCEANS.2006.307050} {\emph {\bibinfo {booktitle} {Oceans
  2006}}}\ (\bibinfo {year} {2006})\ pp.\ \bibinfo {pages} {1--4}\BibitemShut
  {NoStop}%
\bibitem [{\citenamefont {Watson}\ \emph {et~al.}(2012)\citenamefont {Watson},
  \citenamefont {Crutchley},\ and\ \citenamefont {Green}}]{Watson_2011}%
  \BibitemOpen
  \bibfield  {author} {\bibinfo {author} {\bibfnamefont {S.~A.}\ \bibnamefont
  {Watson}}, \bibinfo {author} {\bibfnamefont {D.~J.~P.}\ \bibnamefont
  {Crutchley}}, \ and\ \bibinfo {author} {\bibfnamefont {P.~N.}\ \bibnamefont
  {Green}},\ }\bibfield  {title} {\enquote {\bibinfo {title} {The mechatronic
  design of a micro-autonomous underwater vehicle ($\mu$auv)},}\ }\href
  {\doibase 10.1504/IJMA.2012.049397} {\bibfield  {journal} {\bibinfo
  {journal} {Int. J. of Mechatronics and Automation}\ }\textbf {\bibinfo
  {volume} {2 (3)}} (\bibinfo {year} {2012})}\BibitemShut {NoStop}%
\bibitem [{\citenamefont {Jaffe}\ \emph {et~al.}(2017)\citenamefont {Jaffe},
  \citenamefont {Franks}, \citenamefont {Roberts}, \citenamefont {Mirza},
  \citenamefont {Schurgers}, \citenamefont {Kastner},\ and\ \citenamefont
  {Boch}}]{Jaffe2017}%
  \BibitemOpen
  \bibfield  {author} {\bibinfo {author} {\bibfnamefont {J.~S.}\ \bibnamefont
  {Jaffe}}, \bibinfo {author} {\bibfnamefont {P.~J.~S.}\ \bibnamefont
  {Franks}}, \bibinfo {author} {\bibfnamefont {P.~L.~D.}\ \bibnamefont
  {Roberts}}, \bibinfo {author} {\bibfnamefont {D.}~\bibnamefont {Mirza}},
  \bibinfo {author} {\bibfnamefont {C.}~\bibnamefont {Schurgers}}, \bibinfo
  {author} {\bibfnamefont {R.}~\bibnamefont {Kastner}}, \ and\ \bibinfo
  {author} {\bibfnamefont {A.}~\bibnamefont {Boch}},\ }\bibfield  {title}
  {\enquote {\bibinfo {title} {A swarm of autonomous miniature underwater robot
  drifters for exploring submesoscale ocean dynamics},}\ }\href
  {http://dx.doi.org/10.1038/ncomms14189} {\bibfield  {journal} {\bibinfo
  {journal} {Nature Comm.}\ }\textbf {\bibinfo {volume} {8}},\ \bibinfo {pages}
  {14189 EP --} (\bibinfo {year} {2017})}\BibitemShut {NoStop}%
\bibitem [{\citenamefont {Osborn}\ and\ \citenamefont
  {Lueck}(1985)}]{Osborn1985}%
  \BibitemOpen
  \bibfield  {author} {\bibinfo {author} {\bibfnamefont {T.~R.}\ \bibnamefont
  {Osborn}}\ and\ \bibinfo {author} {\bibfnamefont {R.~G.}\ \bibnamefont
  {Lueck}},\ }\bibfield  {title} {\enquote {\bibinfo {title} {Turbulence
  measurements from a towed body},}\ }\href
  {https://doi.org/10.1175/1520-0426(1985)002<0517:TMFATB>2.0.CO;2} {\bibfield
  {journal} {\bibinfo  {journal} {J. Atmospheric and Oceanic Technology}\
  }\textbf {\bibinfo {volume} {2}},\ \bibinfo {pages} {517--527} (\bibinfo
  {year} {1985})}\BibitemShut {NoStop}%
\bibitem [{\citenamefont {Shan}(1997)}]{PhysRevE.55.2780}%
  \BibitemOpen
  \bibfield  {author} {\bibinfo {author} {\bibfnamefont {X.}~\bibnamefont
  {Shan}},\ }\bibfield  {title} {\enquote {\bibinfo {title} {Simulation of
  {R}ayleigh-{B}\'enard convection using a Lattice Boltzmann method},}\ }\href
  {\doibase 10.1103/PhysRevE.55.2780} {\bibfield  {journal} {\bibinfo
  {journal} {Phys. Rev. E}\ }\textbf {\bibinfo {volume} {55}},\ \bibinfo
  {pages} {2780--2788} (\bibinfo {year} {1997})}\BibitemShut {NoStop}%
\bibitem [{\citenamefont {Perlekar}\ \emph {et~al.}(2012)\citenamefont
  {Perlekar}, \citenamefont {Biferale}, \citenamefont {Sbragaglia},
  \citenamefont {Srivastava},\ and\ \citenamefont
  {Toschi}}]{doi:10.1063/1.4719144}%
  \BibitemOpen
  \bibfield  {author} {\bibinfo {author} {\bibfnamefont {P.}~\bibnamefont
  {Perlekar}}, \bibinfo {author} {\bibfnamefont {L.}~\bibnamefont {Biferale}},
  \bibinfo {author} {\bibfnamefont {M.}~\bibnamefont {Sbragaglia}}, \bibinfo
  {author} {\bibfnamefont {S.}~\bibnamefont {Srivastava}}, \ and\ \bibinfo
  {author} {\bibfnamefont {F.}~\bibnamefont {Toschi}},\ }\bibfield  {title}
  {\enquote {\bibinfo {title} {Droplet size distribution in homogeneous
  isotropic turbulence},}\ }\href {https://doi.org/10.1063/1.4719144}
  {\bibfield  {journal} {\bibinfo  {journal} {Phys. Fluids}\ }\textbf {\bibinfo
  {volume} {24}},\ \bibinfo {pages} {065101} (\bibinfo {year}
  {2012})}\BibitemShut {NoStop}%
\bibitem [{\citenamefont {Mathai}\ \emph {et~al.}(2016)\citenamefont {Mathai},
  \citenamefont {Calzavarini}, \citenamefont {Brons}, \citenamefont {Sun},\
  and\ \citenamefont {Lohse}}]{Mathai_2016}%
  \BibitemOpen
  \bibfield  {author} {\bibinfo {author} {\bibfnamefont {V.}~\bibnamefont
  {Mathai}}, \bibinfo {author} {\bibfnamefont {E.}~\bibnamefont {Calzavarini}},
  \bibinfo {author} {\bibfnamefont {J.}~\bibnamefont {Brons}}, \bibinfo
  {author} {\bibfnamefont {C.}~\bibnamefont {Sun}}, \ and\ \bibinfo {author}
  {\bibfnamefont {D.}~\bibnamefont {Lohse}},\ }\bibfield  {title} {\enquote
  {\bibinfo {title} {Microbubbles and microparticles are not faithful tracers
  of turbulent acceleration},}\ }\href@noop {} {\bibfield  {journal} {\bibinfo
  {journal} {Phys. Rev. Lett.}\ } (\bibinfo {year} {2016})}\BibitemShut
  {NoStop}%
\bibitem [{\citenamefont {Voth}\ \emph {et~al.}(2002)\citenamefont {Voth},
  \citenamefont {La~Porta}, \citenamefont {Crawford}, \citenamefont
  {Alexander},\ and\ \citenamefont {Bodenschatz}}]{voth2002}%
  \BibitemOpen
  \bibfield  {author} {\bibinfo {author} {\bibfnamefont {G.~A.}\ \bibnamefont
  {Voth}}, \bibinfo {author} {\bibfnamefont {A.}~\bibnamefont {La~Porta}},
  \bibinfo {author} {\bibfnamefont {A.~M.}\ \bibnamefont {Crawford}}, \bibinfo
  {author} {\bibfnamefont {J.}~\bibnamefont {Alexander}}, \ and\ \bibinfo
  {author} {\bibfnamefont {E.}~\bibnamefont {Bodenschatz}},\ }\bibfield
  {title} {\enquote {\bibinfo {title} {Measurement of particle accelerations in
  fully developed turbulence},}\ }\href {\doibase 10.1017/S0022112002001842}
  {\bibfield  {journal} {\bibinfo  {journal} {J. Fluid Mech.}\ }\textbf
  {\bibinfo {volume} {469}},\ \bibinfo {pages} {121--160} (\bibinfo {year}
  {2002})}\BibitemShut {NoStop}%
\bibitem [{\citenamefont {Zbrozek}(1958)}]{Zbrozek_1958}%
  \BibitemOpen
  \bibfield  {author} {\bibinfo {author} {\bibfnamefont {J.~K.}\ \bibnamefont
  {Zbrozek}},\ }\bibfield  {title} {\enquote {\bibinfo {title} {Some effects of
  atmospheric turbulence on aircraft},}\ }\href@noop {} {\bibfield  {journal}
  {\bibinfo  {journal} {Weather}\ }\textbf {\bibinfo {volume} {13}},\ \bibinfo
  {pages} {215-- 227} (\bibinfo {year} {1958})}\BibitemShut {NoStop}%
\bibitem [{\citenamefont {Warhaft}(2000)}]{Warhaft_2000}%
  \BibitemOpen
  \bibfield  {author} {\bibinfo {author} {\bibfnamefont {Z.}~\bibnamefont
  {Warhaft}},\ }\bibfield  {title} {\enquote {\bibinfo {title} {Passive scalars
  in turbulent flows},}\ }\href@noop {} {\bibfield  {journal} {\bibinfo
  {journal} {Ann. Rev. Fluid Mech.}\ }\textbf {\bibinfo {volume} {32}},\
  \bibinfo {pages} {203--240} (\bibinfo {year} {2000})}\BibitemShut {NoStop}%
\bibitem [{\citenamefont {Biferale}\ \emph {et~al.}(2011)\citenamefont
  {Biferale}, \citenamefont {Calzavarini},\ and\ \citenamefont
  {Toschi}}]{Biferale_2011}%
  \BibitemOpen
  \bibfield  {author} {\bibinfo {author} {\bibfnamefont {L.}~\bibnamefont
  {Biferale}}, \bibinfo {author} {\bibfnamefont {E.}~\bibnamefont
  {Calzavarini}}, \ and\ \bibinfo {author} {\bibfnamefont {F.}~\bibnamefont
  {Toschi}},\ }\bibfield  {title} {\enquote {\bibinfo {title} {Multi-time
  multi-scale correlation functions in hydrodynamic turbulence},}\ }\href@noop
  {} {\bibfield  {journal} {\bibinfo  {journal} {Phys. Fluids}\ }\textbf
  {\bibinfo {volume} {23}},\ \bibinfo {pages} {085107} (\bibinfo {year}
  {2011})}\BibitemShut {NoStop}%
\bibitem [{\citenamefont {Tennekes}\ and\ \citenamefont
  {Lumley}(1972)}]{tennekes1972first}%
  \BibitemOpen
  \bibfield  {author} {\bibinfo {author} {\bibfnamefont {H.}~\bibnamefont
  {Tennekes}}\ and\ \bibinfo {author} {\bibfnamefont {J.L.}\ \bibnamefont
  {Lumley}},\ }\href {https://books.google.fr/books?id=h4coCj-lN0cC} {\emph
  {\bibinfo {title} {A First Course in Turbulence}}},\ A First Course in
  Turbulence\ (\bibinfo  {publisher} {MIT Press},\ \bibinfo {year}
  {1972})\BibitemShut {NoStop}%
\bibitem [{\citenamefont {Schmitt}\ and\ \citenamefont
  {Huang}(2016)}]{sh2016book}%
  \BibitemOpen
  \bibfield  {author} {\bibinfo {author} {\bibfnamefont {F.~G.}\ \bibnamefont
  {Schmitt}}\ and\ \bibinfo {author} {\bibfnamefont {Y.}~\bibnamefont
  {Huang}},\ }\href@noop {} {\emph {\bibinfo {title} {Stochastic Analysis of
  Scaling Time Series: From Turbulence Theory to Applications}}}\ (\bibinfo
  {publisher} {Cambridge Univ. Press},\ \bibinfo {year} {2016})\BibitemShut
  {NoStop}%
\bibitem [{\citenamefont {Sreenivasan}(1996)}]{Sreenivasan96}%
  \BibitemOpen
  \bibfield  {author} {\bibinfo {author} {\bibfnamefont {Katepalli~R.}\
  \bibnamefont {Sreenivasan}},\ }\bibfield  {title} {\enquote {\bibinfo {title}
  {The passive scalar spectrum and the Obukhov-Corrsin constant},}\ }\href@noop
  {} {\bibfield  {journal} {\bibinfo  {journal} {Physics of Fluids}\ }\textbf
  {\bibinfo {volume} {8}},\ \bibinfo {pages} {189--196} (\bibinfo {year}
  {1996})}\BibitemShut {NoStop}%
\bibitem [{\citenamefont {Huang}\ \emph {et~al.}(2013)\citenamefont {Huang},
  \citenamefont {Biferale}, \citenamefont {Calzavarini}, \citenamefont {Sun},\
  and\ \citenamefont {Toschi}}]{PhysRevE.87.041003}%
  \BibitemOpen
  \bibfield  {author} {\bibinfo {author} {\bibfnamefont {Y.}~\bibnamefont
  {Huang}}, \bibinfo {author} {\bibfnamefont {L.}~\bibnamefont {Biferale}},
  \bibinfo {author} {\bibfnamefont {E.}~\bibnamefont {Calzavarini}}, \bibinfo
  {author} {\bibfnamefont {C.}~\bibnamefont {Sun}}, \ and\ \bibinfo {author}
  {\bibfnamefont {F.}~\bibnamefont {Toschi}},\ }\bibfield  {title} {\enquote
  {\bibinfo {title} {Lagrangian single-particle turbulent statistics through
  the Hilbert-Huang transform},}\ }\href {\doibase 10.1103/PhysRevE.87.041003}
  {\bibfield  {journal} {\bibinfo  {journal} {Phys. Rev. E}\ }\textbf {\bibinfo
  {volume} {87}},\ \bibinfo {pages} {041003} (\bibinfo {year}
  {2013})}\BibitemShut {NoStop}%
\bibitem [{\citenamefont {Watmuff}\ \emph {et~al.}(1983)\citenamefont
  {Watmuff}, \citenamefont {Perry},\ and\ \citenamefont
  {Chong}}]{Watmuff_1983}%
  \BibitemOpen
  \bibfield  {author} {\bibinfo {author} {\bibfnamefont {J.~H.}\ \bibnamefont
  {Watmuff}}, \bibinfo {author} {\bibfnamefont {A.~E.}\ \bibnamefont {Perry}},
  \ and\ \bibinfo {author} {\bibfnamefont {M.~S.}\ \bibnamefont {Chong}},\
  }\bibfield  {title} {\enquote {\bibinfo {title} {A flying hot-wire system},}\
  }\href@noop {} {\bibfield  {journal} {\bibinfo  {journal} {Exp. Fluids}\
  }\textbf {\bibinfo {volume} {1}},\ \bibinfo {pages} {63--71} (\bibinfo {year}
  {1983})}\BibitemShut {NoStop}%
\bibitem [{\citenamefont {Pinton}\ and\ \citenamefont
  {Labb\'e}(1994)}]{Pinton_1994}%
  \BibitemOpen
  \bibfield  {author} {\bibinfo {author} {\bibfnamefont {J.-F.}\ \bibnamefont
  {Pinton}}\ and\ \bibinfo {author} {\bibfnamefont {R.}~\bibnamefont
  {Labb\'e}},\ }\bibfield  {title} {\enquote {\bibinfo {title} {Correction to
  the taylor hypothesis in swirling flows},}\ }\href@noop {} {\bibfield
  {journal} {\bibinfo  {journal} {J. de Physique II}\ }\textbf {\bibinfo
  {volume} {4 (9)}},\ \bibinfo {pages} {1461--1468} (\bibinfo {year}
  {1994})}\BibitemShut {NoStop}%
\bibitem [{\citenamefont {Yeung}(2001)}]{yeung_2001}%
  \BibitemOpen
  \bibfield  {author} {\bibinfo {author} {\bibfnamefont {P.~K.}\ \bibnamefont
  {Yeung}},\ }\bibfield  {title} {\enquote {\bibinfo {title} {Lagrangian
  characteristics of turbulence and scalar transport in Direct Numerical
  Simulations},}\ }\href {\doibase 10.1017/S0022112000002391} {\bibfield
  {journal} {\bibinfo  {journal} {J. Fluid Mech.}\ }\textbf {\bibinfo {volume}
  {427}},\ \bibinfo {pages} {241-274} (\bibinfo {year} {2001})}\BibitemShut
  {NoStop}%
\bibitem [{\citenamefont {Bec}\ \emph {et~al.}(2014)\citenamefont {Bec},
  \citenamefont {Homann},\ and\ \citenamefont
  {Krstulovic}}]{PhysRevLett.112.234503}%
  \BibitemOpen
  \bibfield  {author} {\bibinfo {author} {\bibfnamefont {J.}~\bibnamefont
  {Bec}}, \bibinfo {author} {\bibfnamefont {H.}~\bibnamefont {Homann}}, \ and\
  \bibinfo {author} {\bibfnamefont {G.}~\bibnamefont {Krstulovic}},\ }\bibfield
   {title} {\enquote {\bibinfo {title} {Clustering, fronts, and heat transfer
  in turbulent suspensions of heavy particles},}\ }\href {\doibase
  10.1103/PhysRevLett.112.234503} {\bibfield  {journal} {\bibinfo  {journal}
  {Phys. Rev. Lett.}\ }\textbf {\bibinfo {volume} {112}},\ \bibinfo {pages}
  {234503} (\bibinfo {year} {2014})}\BibitemShut {NoStop}%
\bibitem [{\citenamefont {Arn\'eodo}\ \emph {et~al.}(2008)\citenamefont
  {Arn\'eodo}, \citenamefont {Benzi}, \citenamefont {Berg}, \citenamefont
  {Biferale},\ and\ \citenamefont {Bodenschatz~et al.}}]{Arneodo_2008}%
  \BibitemOpen
  \bibfield  {author} {\bibinfo {author} {\bibfnamefont {A.}~\bibnamefont
  {Arn\'eodo}}, \bibinfo {author} {\bibfnamefont {R.}~\bibnamefont {Benzi}},
  \bibinfo {author} {\bibfnamefont {J.}~\bibnamefont {Berg}}, \bibinfo {author}
  {\bibfnamefont {L.}~\bibnamefont {Biferale}}, \ and\ \bibinfo {author}
  {\bibfnamefont {E.}~\bibnamefont {Bodenschatz~et al.}},\ }\bibfield  {title}
  {\enquote {\bibinfo {title} {Universal intermittent properties of particle
  trajectories in highly turbulent flows},}\ }\href@noop {} {\bibfield
  {journal} {\bibinfo  {journal} {Phys. Rev. Lett.}\ }\textbf {\bibinfo
  {volume} {100}},\ \bibinfo {pages} {254504} (\bibinfo {year}
  {2008})}\BibitemShut {NoStop}%
\bibitem [{\citenamefont {Biferale}\ \emph {et~al.}(2005)\citenamefont
  {Biferale}, \citenamefont {Boffetta}, \citenamefont {Celani}, \citenamefont
  {Lanotte},\ and\ \citenamefont {Toschi}}]{Biferale2005}%
  \BibitemOpen
  \bibfield  {author} {\bibinfo {author} {\bibfnamefont {L.}~\bibnamefont
  {Biferale}}, \bibinfo {author} {\bibfnamefont {G.}~\bibnamefont {Boffetta}},
  \bibinfo {author} {\bibfnamefont {A.}~\bibnamefont {Celani}}, \bibinfo
  {author} {\bibfnamefont {A.}~\bibnamefont {Lanotte}}, \ and\ \bibinfo
  {author} {\bibfnamefont {F.}~\bibnamefont {Toschi}},\ }\bibfield  {title}
  {\enquote {\bibinfo {title} {Particle trapping in three-dimensional fully
  developed turbulence},}\ }\href {\doibase 10.1063/1.1846771} {\bibfield
  {journal} {\bibinfo  {journal} {Phys. Fluids}\ }\textbf {\bibinfo {volume}
  {17}},\ \bibinfo {pages} {021701} (\bibinfo {year} {2005})} \BibitemShut {NoStop}%
\bibitem [{\citenamefont {Buzzicotti}\ \emph {et~al.}(2016)\citenamefont
  {Buzzicotti}, \citenamefont {Bhatnagar}, \citenamefont {Biferale},
  \citenamefont {Lanotte},\ and\ \citenamefont {Ray}}]{Buzzicotti2016}%
  \BibitemOpen
  \bibfield  {author} {\bibinfo {author} {\bibfnamefont {M.}~\bibnamefont
  {Buzzicotti}}, \bibinfo {author} {\bibfnamefont {A.}~\bibnamefont
  {Bhatnagar}}, \bibinfo {author} {\bibfnamefont {L.}~\bibnamefont {Biferale}},
  \bibinfo {author} {\bibfnamefont {A.~S.}\ \bibnamefont {Lanotte}}, \ and\
  \bibinfo {author} {\bibfnamefont {S.~S.}\ \bibnamefont {Ray}},\ }\bibfield
  {title} {\enquote {\bibinfo {title} {Lagrangian statistics for
  Navier-Stokes turbulence under Fourier-mode reduction: fractal and
  homogeneous decimations},}\ }\href
  {http://stacks.iop.org/1367-2630/18/i=11/a=113047} {\bibfield  {journal}
  {\bibinfo  {journal} {New J. Phys.}\ }\textbf {\bibinfo {volume} {18}},\
  \bibinfo {pages} {113047} (\bibinfo {year} {2016})}\BibitemShut {NoStop}%
\bibitem [{\citenamefont {Sreenivasan}\ and\ \citenamefont
  {Schumacher}(2010)}]{Sreenivasan1561}%
  \BibitemOpen
  \bibfield  {author} {\bibinfo {author} {\bibfnamefont {K.~R.}\ \bibnamefont
  {Sreenivasan}}\ and\ \bibinfo {author} {\bibfnamefont {J.}~\bibnamefont
  {Schumacher}},\ }\bibfield  {title} {\enquote {\bibinfo {title} {Lagrangian
  views on turbulent mixing of passive scalars},}\ }\href {\doibase
  10.1098/rsta.2009.0140} {\bibfield  {journal} {\bibinfo  {journal} {Phil.
  Trans. Royal Society}\ }\textbf {\bibinfo {volume} {368}},\ \bibinfo {pages}
  {1561--1577} (\bibinfo {year} {2010})}
  \BibitemShut {NoStop}%
\bibitem [{\citenamefont {Jim\'enez}(1997)}]{Jimenez_1997}%
  \BibitemOpen
  \bibfield  {author} {\bibinfo {author} {\bibfnamefont {J.}~\bibnamefont
  {Jim\'enez}},\ }\bibfield  {title} {\enquote {\bibinfo {title} {Oceanic
  turbulence at millimeter scales},}\ }\href@noop {} {\bibfield  {journal}
  {\bibinfo  {journal} {Scientia Marina}\ }\textbf {\bibinfo {volume} {61}},\
  \bibinfo {pages} {47 -- 56} (\bibinfo {year} {1997})}\BibitemShut {NoStop}%
\bibitem [{\citenamefont {Thorpe}(2007)}]{Thorpe2007book}%
  \BibitemOpen
  \bibfield  {author} {\bibinfo {author} {\bibfnamefont {S.~A.}\ \bibnamefont
  {Thorpe}},\ }\href@noop {} {\emph {\bibinfo {title} {An Introduction to Ocean
  Turbulence}}}\ (\bibinfo  {publisher} {Cambridge Univ. Press},\ \bibinfo
  {year} {2007})\BibitemShut {NoStop}%
\bibitem [{\citenamefont {Korotenko}\ \emph {et~al.}(2013)\citenamefont
  {Korotenko}, \citenamefont {Sentchev}, \citenamefont {Schmitt},\ and\
  \citenamefont {Jouanneau}}]{KOROTENKO201321}%
  \BibitemOpen
  \bibfield  {author} {\bibinfo {author} {\bibfnamefont {K.}~\bibnamefont
  {Korotenko}}, \bibinfo {author} {\bibfnamefont {A.}~\bibnamefont {Sentchev}},
  \bibinfo {author} {\bibfnamefont {F.~G.}\ \bibnamefont {Schmitt}}, \ and\
  \bibinfo {author} {\bibfnamefont {N.}~\bibnamefont {Jouanneau}},\ }\bibfield
  {title} {\enquote {\bibinfo {title} {Variability of turbulent quantities in
  the tidal bottom boundary layer: Case study in the eastern english
  channel},}\ }\href {\doibase https://doi.org/10.1016/j.csr.2013.03.001}
  {\bibfield  {journal} {\bibinfo  {journal} {Continental Shelf Research}\
  }\textbf {\bibinfo {volume} {58}},\ \bibinfo {pages} {21 -- 31} (\bibinfo
  {year} {2013})}\BibitemShut {NoStop}%
\bibitem [{\citenamefont {Torney}\ and\ \citenamefont
  {Neufeld}(2008)}]{Torney_2008}%
  \BibitemOpen
  \bibfield  {author} {\bibinfo {author} {\bibfnamefont {C.}~\bibnamefont
  {Torney}}\ and\ \bibinfo {author} {\bibfnamefont {Z.}~\bibnamefont
  {Neufeld}},\ }\bibfield  {title} {\enquote {\bibinfo {title} {Phototactic
  clustering of swimming microorganisms in a turbulent velocity field},}\
  }\href@noop {} {\bibfield  {journal} {\bibinfo  {journal} {Phys. Rev. Lett.}\
  }\textbf {\bibinfo {volume} {101}},\ \bibinfo {pages} {078105} (\bibinfo
  {year} {2008})}\BibitemShut {NoStop}%
\bibitem [{\citenamefont {Dervaux}\ \emph {et~al.}(2017)\citenamefont
  {Dervaux}, \citenamefont {Capellazzi~Resta},\ and\ \citenamefont
  {Brunet}}]{Dervaux_2017}%
  \BibitemOpen
  \bibfield  {author} {\bibinfo {author} {\bibfnamefont {J.}~\bibnamefont
  {Dervaux}}, \bibinfo {author} {\bibfnamefont {M.}~\bibnamefont
  {Capellazzi~Resta}}, \ and\ \bibinfo {author} {\bibfnamefont
  {P.}~\bibnamefont {Brunet}},\ }\bibfield  {title} {\enquote {\bibinfo {title}
  {Light-controlled flows in active fluids},}\ }\href@noop {} {\bibfield
  {journal} {\bibinfo  {journal} {Nature Phys.}\ }\textbf {\bibinfo {volume}
  {13}} (\bibinfo {year} {2017})}\BibitemShut {NoStop}%
\end{thebibliography}

%

\end{document}


\title{
Supplementary Material for:\\
\textit{``Propelled micro-probes in turbulence''}
}

\author{\textsc{E. Calzavarini}}\email[]{enrico.calzavarini@polytech-lille.fr} 
\affiliation{ \textit{Univ. Lille, CNRS, FRE 3723, LML, Laboratoire de M\'{e}canique de Lille, F 59000 Lille, France}}
\author{\textsc{Y. X. Huang}}\email[]{yongxianghuang@xmu.edu.cn} 
\affiliation{ \textit{State Key Laboratory of Marine Environmental Science, College of Ocean and Earth Sciences, Xiamen University,
Xiamen 361102, People's Republic of China}}
\author{\textsc{F. G. Schmitt}}\email[]{francois.schmitt@cnrs.fr} 
\affiliation{ \textit{Univ. Lille, CNRS, Univ. Littoral Cote d'Opale, UMR 8187, LOG,Laboratoire d'Oc\'{e}anologie et de G\'{e}oscience, F 62930 Wimereux, France}}
\author{\textsc{L. P. Wang}}\email[]{lipo.wang@sjtu.edu.cn} 
\affiliation{ \textit{UM-SJTU Joint Institute, Shanghai JiaoTong University, Shanghai, 200240, People's Republic of China}}


\date{\today}

\widetext
\begin{center}
\textbf{\large Supplementary Material for:\\}
\vspace{3mm} 
\textbf{\large \textit{``Propelled micro-probes in turbulence''}}\\
\vspace{3mm} 
{\small
\textsc{E. Calzavarini$^1$, Y. X. Huang$^2$, F. G. Schmitt$^3$, L. P. Wang$^4$}\\
$^1$\textit{Univ. Lille, CNRS, FRE 3723, LML, Laboratoire de M\'{e}canique de Lille, F 59000 Lille, France}\\
$^2$\textit{State Key Laboratory of Marine Environmental Science, College of Ocean and Earth Sciences, Xiamen University, Xiamen 361102, People's Republic of China}\\
$^3$\textit{Univ. Lille, CNRS, Univ. Littoral Cote d'Opale, UMR 8187, LOG,Laboratoire d'Oc\'{e}anologie et de G\'{e}oscience, F 62930 Wimereux, France}\\
$^4$\textit{UM-SJTU Joint Institute, Shanghai JiaoTong University, Shanghai, 200240, People's Republic of China}\\
}
\vspace{1mm}
\today

\end{center}
\setcounter{equation}{0}
\setcounter{figure}{0}
\setcounter{table}{0}
\setcounter{page}{1}
\setcounter{section}{0}
\setcounter{subsection}{0}
\makeatletter
\renewcommand{\theequation}{S\arabic{equation}}
\renewcommand{\thefigure}{S\arabic{figure}}
\renewcommand{\thetable}{S\arabic{table}}

\section{Single-point statistical relations for fluid velocity and scalar}

\subsection{Variance of time derivatives}
We consider the equation of motion for the probe, $\dot{\bm{x}}_s = \bm{u}( {\bm{x}}_s(t), t)    + \bm{v}_s $,
and compute its total time derivative with respect to time, this leads to:
\begin{equation}\label{sacc}
\ddot{\bm{x}}_s = \partial_t \bm{u}( {\bm{x}}_s(t), t)  +  \dot{\bm{x}}_s  \cdot \bm{\partial} \bm{u}( {\bm{x}}_s(t), t)  = D_t \bm{u}( {\bm{x}}_s(t), t)    + \bm{v}_s \cdot \bm{\partial} \bm{u}( {\bm{x}}_s(t), t) 
\end{equation}
where $D_t (\bullet)  = \partial_t(\bullet) + \bm{u} \cdot \bm{\partial} (\bullet)$ is the material fluid derivative. 
We now compute the variance for a single cartesian component  of the equation and assume that $D_t u_i$ and $\bm{v}_s \cdot \bm{\partial} u_i$ are statistically independent, this leads to:
\begin{equation}\label{temp_derivative}
\langle \ddot{\bm{x}}_{s,i}^2 \rangle = \langle \left( D_t  u_i( {\bm{x}}_s(t), t) \right)^2 \rangle    +  \langle ( \bm{v}_s \cdot \bm{\partial} u_i( {\bm{x}}_s(t), t) )^2 \rangle. 
\end{equation}
Note that because the probes do not cluster and explore the space homogeneously one can substitute ${\bm{x}}_s(t)$ one can also write:
\begin{equation}\label{temp_derivative2}
\langle \ddot{\bm{x}}_{s,i}^2 \rangle = \langle \left( D_t  u_i( {\bm{x}}, t) \right)^2 \rangle    +  \langle ( \bm{v}_s \cdot \bm{\partial} u_i( \bm{x}, t) )^2 \rangle,
\end{equation}
where  ${\bm{x}}$ is a generic spatial coordinate, meaning that the average is taken over all position (volume average) or equivalently over a set of homogeneously distributed points (as for instance an ensemble of  Lagrangian tracers  $\bm{x}_L(t)$).
In statistical isotropic turbulent conditions the following relations hold:
$$\langle \left( D_t  u_i( {\bm{x}}, t) \right)^2 \rangle = a_0\ \epsilon/ \tau_{\eta}, \quad  \langle \left( \partial_j u_i( {\bm{x}}, t) \right)^2 \rangle = 2\ \langle \left( \partial_i  u_i( {\bm{x}}, t) \right)^2 \rangle = 2/15\ \epsilon/ \tau_{\eta}.$$
The first relation can be taken as a definition of the the Heisenberg-Yaglom constant $a_0$, while the seconds are kinematic relations which take into account also the incompressibility constraint (see Hinze book for a derivation).
The second term  in (\ref{temp_derivative}) with the additional assumption of isotropy of the probe ensemble can be rewritten as $\frac{v_s^2}{3}  \Sigma_j \langle  (\partial_j u_i)^2  \rangle = \frac{v_s^2}{9} \epsilon/ \tau_{\eta}$. This altogether leads to equation (5) of the paper.\\

The case of the time derivative of the scalar measured by a probe $ \theta_s =\theta( {\bm{x}}_s(t), t)$ can be treated in a similar way. The following relation holds: 
\begin{equation}\label{temp_derivative}
\dot{\theta}_s = D_t \theta( {\bm{x}}_s(t), t)    + \bm{v}_s \cdot \bm{\partial} \theta( {\bm{x}}_s(t), t) 
\end{equation}
As before we compute the variance of the above quantity and assume that $D_t \theta$ and $\bm{v}_s \cdot \bm{\partial} \theta$ are statistically independent, this leads to:
\begin{equation}\label{temp_derivative_var}
\langle \dot{\theta}_s^2 \rangle = \langle \left( D_t \theta( {\bm{x}}, t) \right)^2 \rangle    +  \langle ( \bm{v}_s \cdot \bm{\partial} \theta( {\bm{x}}, t) )^2 \rangle 
\end{equation}
The first term can be  rewritten in terms of  a new quantity $b_0$, equivalent of the HY constant  for fluid acceleration defined as $ \langle \dot{\theta}_L^2 \rangle = b_0\ \epsilon_{\theta}/ \tau_{\eta}$ and $\dot{\theta}_L = D_t \theta(\bm{x}_L(t), t)$ is the time derivative of scalar along a fluid tracer and $\epsilon_{\theta} = \kappa \langle  |\partial \theta|^2 \rangle$.

The second term on the right-hand-side of (\ref{temp_derivative_var}), can be rewritten as $\frac{v_s^2}{3}  \langle |\partial \theta|^2 \rangle  $ with the assumption of isotropy of scalar field and probe set. By combining these relations we find the prediction give in equation (6) of the paper. A relation that we could check only for the case $Sc=1$.

\subsection{Flatness of time derivatives}
The above relations can be extended to the flatness factor both for the velocity and the scalar. Under the same hypothesis, the fourth order moment relations for velocity and scalar derivatives are, 
\begin{equation}\label{temp_derivative_m4_v}
\langle \ddot{x}_{s,i}^4 \rangle = \langle \left( D_t u_i( {\bm{x}}_s(t), t) \right)^4 \rangle    +  6  \langle \left( D_t u_i( {\bm{x}}_s(t), t) \right)^2 \rangle \langle ( \bm{v}_s \cdot \partial u_i( {\bm{x}}_s(t), t) )^2 \rangle + \langle ( \bm{v}_s \cdot \partial u_i( {\bm{x}}_s(t), t) )^4 \rangle, 
\end{equation}
\begin{equation}\label{temp_derivative_m4}
\langle \dot{\theta}_s^4 \rangle = \langle \left( D_t \theta( {\bm{x}}_s(t), t) \right)^4 \rangle    +  6  \langle \left( D_t \theta( {\bm{x}}_s(t), t) \right)^2 \rangle \langle ( \bm{v}_s \cdot \partial \theta( {\bm{x}}_s(t), t) )^2 \rangle + \langle ( \bm{v}_s \cdot \partial \theta( {\bm{x}}_s(t), t) )^4 \rangle. 
\end{equation}
Again we take into account the isotropy of the Eulerian fields together with the isotropy of the set of probes:

$$ \langle ( \bm{v}_s \cdot \partial u_i( {\bm{x}}_s(t), t) )^4 \rangle =  \frac{v_s^4}{3} \Sigma_j \langle(\partial_j u_i )^4 \rangle =  \frac{v_s^4}{3} \left( \langle (\partial_{\parallel} u_i )^4 \rangle + 2  \langle (\partial_{\perp} u_i )^4 \rangle \right) $$
$$ \langle ( \bm{v}_s \cdot \partial \theta( {\bm{x}}_s(t), t) )^4 \rangle = \frac{v_s^4}{3}\Sigma_j \langle (\partial_j \theta)^4 \rangle $$
This lead to the following relations for the flatness factors

\begin{equation}\label{temp_derivative_flatness}
\mathcal{F}(\ddot{x}_{s,i})= \frac{\langle \ddot{x}_{s,i}^4 \rangle}{\langle \ddot{x}_{s,i}^2 \rangle^2} = \frac{ \mathcal{F}(D_t u_i)  + \frac{6}{9 a_0} \left( \frac{v_s}{u_{\eta}}  \right)^2  + \left( \mathcal{F}(\partial_{\parallel} u_i) + 8 \mathcal{F}(\partial_{\perp} u_i) \right)   \frac{1}{675 a_0^2} \left( \frac{v_s}{u_{\eta}}  \right)^4  }{  1  + \frac{2}{9 a_0} \left( \frac{v_s}{u_{\eta}}  \right)^2  +   \frac{1}{81 a_0^2}  \left( \frac{v_s}{u_{\eta}}  \right)^4   }
\end{equation}
\begin{equation}\label{temp_derivative_flatness}
\mathcal{F}( \dot{\theta}_s)= \frac{\langle \dot{\theta}_s^4 \rangle}{\langle \dot{\theta}_s^2 \rangle^2} = \frac{ \mathcal{F}( D_t \theta)  + \frac{2 Sc}{b_0} \left( \frac{v_s}{u_{\eta}}  \right)^2  + \mathcal{F}(\partial \theta_s)\   \frac{Sc^2}{9 b_0^2} \left( \frac{v_s}{u_{\eta}}  \right)^4  }{  1  + \frac{2 Sc}{3 b_0} \left( \frac{v_s}{u_{\eta}}  \right)^2  +   \frac{Sc^2}{9 b_0^2}  \left( \frac{v_s}{u_{\eta}}  \right)^4   }
\end{equation}
where $\mathcal{F}(\partial_{\parallel} u_i)$, $\mathcal{F}(\partial_{\perp} u_i)$  and $ \mathcal{F}(\partial \theta_s)$ denote respectively the flatness factor of parallel/transverse velocity gradients and of the scalar.

The asymptotic limit, $v_s \to \infty$, of large propulsions gives respectively $\mathcal{F}(\ddot{x}_{s,i}) =  3/25\ \mathcal{F}(\partial_{\parallel} u_i) +   24/25\ \mathcal{F}(\partial_{\perp} u_i)$ and $\mathcal{F}( \dot{\theta}_s) =  \mathcal{F}(\partial \theta_s)$. Note that it is well known that the single point statistical moments of longitudinal and transverse velocity gradients are different. However while for the second moment their relation is analytically know, their relation has not been derived analytically for the fourth order moment, see \cite{PhysRevE.94.023114} for a recent discussion. In our numerics at $Re_{\lambda} = 125$ we find $\langle (\partial_{\perp} u_i) )^4\rangle  \simeq  5.86 \langle (\partial_{\parallel} u_i)^4 \rangle$, a results that is consistent with the numerical findings by \cite{ishihara_kaneda_yokokawa_itakura_uno_2007}.
The latter results implies  $\mathcal{F}(\partial_{\perp} u_i) \simeq 1.47 \mathcal{F}(\partial_{\parallel} u_i) $, from which one get the asymptotic behaviour  $\mathcal{F}(\ddot{x}_{s,i})  \sim  1.53\  \mathcal{F}(\partial_{\parallel} u_i)$.

\section{Direct Numerical Simulations}

\subsection{Flow parameters}\label{parameters}
For completeness and for future reference we report in table \ref{table:euler} the value of all the DNS parameters and the magnitude of all relevant measured turbulence scales in the performed simulations. 

\begin{table}[h]
\begin{center}
\begin{tabular}{ | c | c | c | c | c | c | c | c | c | c|  c | c | c |}
\hline
$Re_{\lambda}$  & $N^3$& $L$ & $T_{tot}$ & $u'$ & $\lambda$ & $\nu$ & $\eta$ &  $\tau_{\eta}$  & $u_{\eta}$ & $\epsilon$   &  $a_0$\\
\hline
\hline
75 &   $256^3$ & 256 & $1.2 \cdot 10^5$ & $2.44 \cdot 10^{-2}$ & 25.6  & $1/120$ &1.5  & 272 &  $5.515  \cdot 10^{-3}$ &$1.143 \cdot 10^{-7}$   & $2.17 \pm 0.01$ \\
\hline
125 & $512^3$ & 512 & $6.4 \cdot 10^4$ & $3.14 \cdot 10^{-2}$& 33.2 & $1/120$ &1.5 & 272 &  $5.515  \cdot 10^{-3}$  &$1.143 \cdot 10^{-7} $  & $2.72 \pm 0.03$\\
\hline
\end{tabular}
\end{center}
\hspace{-4.6cm}
\begin{tabular}{ | c  | c | c | c | c | c | c | c |}
\hline
$Pe_{\lambda_{\theta}}$  & $Sc$ & $\theta'$  & $\lambda_{\theta}$ & $\eta_{\theta}$ &    $\epsilon_{\theta}$  & $\kappa$ & $b_0$\\
\hline
\hline
  43 &  1 &  $3.05 \cdot 10^{-2}$  & 14.6 & 1.5   & $1.143 \cdot 10^{-7}$ & $1/120$  & $0.50 \pm 0.01$ \\
\hline
  68& 1 &    $3.80 \cdot 10^{-2}$ & 18.1  &  1.5    &$1.143 \cdot 10^{-7} $& $1/120$  & $0.48 \pm 0.01$\\
\hline
\end{tabular}
\caption{DNS parameters. All dimensional quantities are given in terms of time ($\delta t$) and space ($\delta x$) discretisation units. 
Furthermore, in the LB algorithm used we assume  $\delta t = \delta x = 1$.  
$Re_{\lambda}$ is the computed Taylor-scale based Reynolds number $Re_{\lambda} = u' \lambda / \nu$, with $u'$  the single-component root-mean-square velocity (note that $u_{rms} = \sqrt{3} u'$),  $\lambda$ is the Taylor micro scale and $\nu$ the kinematic viscosity. $N^3$ is the total number of grid points, $L$ size of the periodic box, $T_{tot}$ is the total duration of the simulation,  $\eta$ , $\tau_{\eta}$  and  $u_{\eta}$ are the Kolmogorov scales, $\epsilon$  kinetic energy dissipation rate,finally $a_0$ is the Heisenberg-Yaglom acceleration constant, which is defined as $a_0  \equiv \langle a_i^2 \rangle \tau_{\eta}^2/\eta$ , where $ \langle a_i^2 \rangle$ denotes the variance of a single cartesian component of the fluid acceleration.
Concerning the scalar:  $\kappa$ is the scalar diffusivity.  $Sc$ is the Schmidt number $Sc=\nu/\kappa$, while the Taylor-scale based P\'eclet number is  defined as $Pe_{\lambda_{\theta}} =  u' \lambda_{\theta}/\kappa$. The root-mean-square fluctuations of the scalar is denoted as $\theta'$, $\lambda_{\theta} = \sqrt{3\ \theta'^2/ \epsilon_\theta} $ is the Taylor scale equivalent for the scalar field, the Batchelor scale is denoted as $\eta_{\theta} = \eta/\sqrt{Sc}$, $\epsilon_{\theta} = \langle \kappa\ \Sigma_i (\partial_i \theta)^2 \rangle$ is the scalar energy dissipation rate.}\label{table:euler}
\end{table}

\subsection{A Lagrangian validation test}\label{validation}

\begin{figure}[!htb]
\begin{center}
 \includegraphics[width = 0.55\columnwidth]{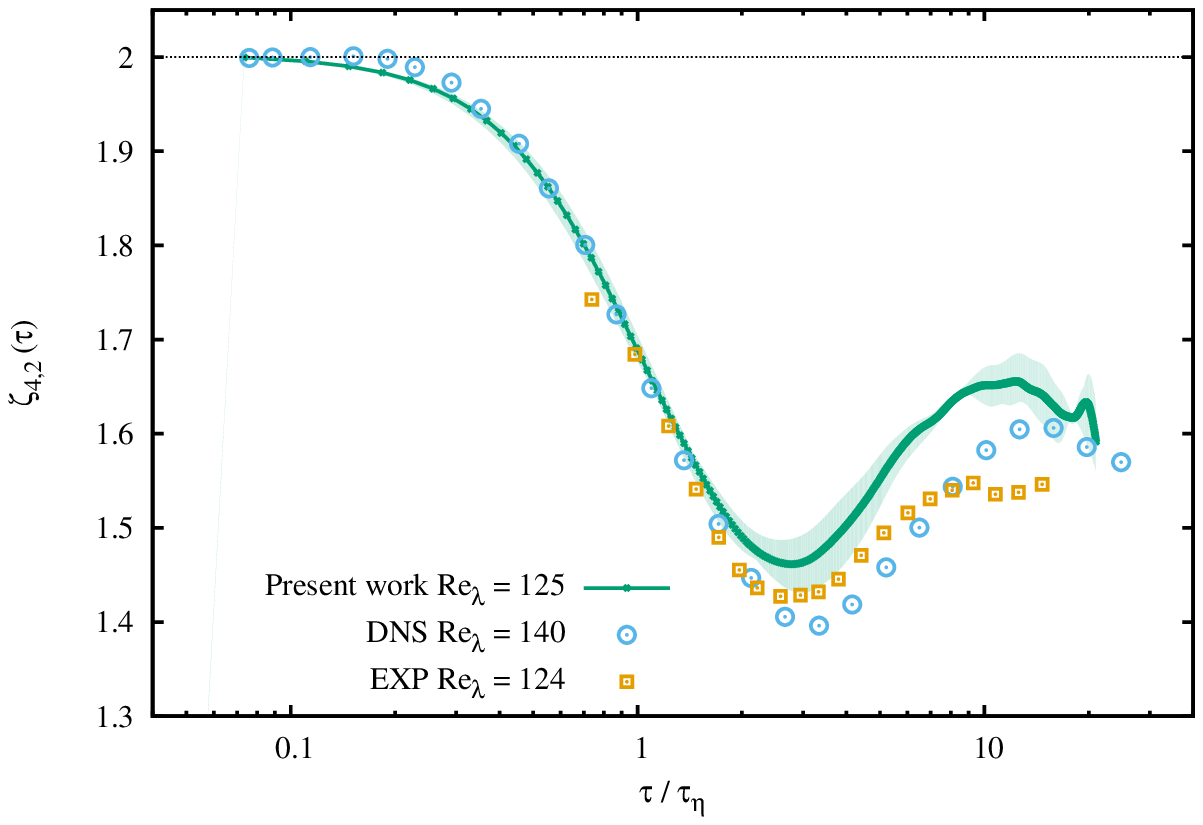}
\caption{Logarithmic derivative of the fourth order moments of Lagrangian fluid velocity time increments versus the second order one as in Eq. (13) of the paper. Comparison of present simulations at $Re_{\lambda} =125$ with simulations and experiments at comparable Reynolds number from Ref. \cite{Arneodo_2008}.}
\label{sfcomparison}
\end{center}
\end{figure}

In order to assess the appropriate resolution of small-scale intermittent fluctuations in the present DNS based on LB method, we compare the local scaling exponent $\zeta_{4,2}(\tau)$ with the one measured in experiment and in simulations at similar $Re_{\lambda}$ number, which were reported in Ref. \cite{Arneodo_2008}. The plot of such a comparison is reported in figure \ref{sfcomparison}, note that the 
experimental data comes from Particle Tracking Velocimetry (PTV) measurements in a turbulent Von Karman flow flow setup at $Re_{\lambda} = 124$ while, the  numerical data  at $Re_{\lambda} = 140$ and were computed by DNS particle tracking in homogeneous isotropic turbulence in a pseudo-spectral simulation in a tri-periodic cubic box.



%
